\documentclass{jfm}

\usepackage{graphicx}
\usepackage{natbib}
\usepackage{subfigure}
\usepackage{amsmath}
\usepackage{amssymb}
\usepackage{graphicx}
\usepackage{epstopdf, epsfig}
\usepackage{xcolor}
\usepackage{multirow}
\usepackage{hyperref}
\usepackage{siunitx}

\newcommand{\ud}{\mathrm{d}}
\shorttitle{Turbulent Newtonian and non-Newtonian fluid flow in rough channels}
\shortauthor{C. Narayanan et al.}

\title[Turbulent flow of Newtonian and non-Newtonian fluids in rough channels.]{Turbulent flow of Newtonian and non-Newtonian fluids in rough channels: Part I: Roughness effect on Newtonian fluids}

\author[C. Narayanan et al.]%
{C.\ls \ls N\ls A\ls R\ls A\ls Y\ls A\ls N\ls A\ls  N\ls$^{1}$, S.\ls \ls N\ls A\ls U\ls E\ls R\ls$^{1}$, J.\ls S.\ls \ls S\ls I\ls N\ls G\ls H\ls$^{1}$, 
R.\ls \ls B\ls E\ls L\ls T\ls$^{2}$, T.\ls \ls P\ls A\ls L\ls E\ls R\ls M\ls O\ls$^{2}$ \ls  \and D.\ls \ls L\ls A\ls K\ls E\ls H\ls A\ls L\ls$^{1}$$^{*}$} 
\affiliation{$^{1}$AFRY AMS, Zürich, Switzerland
$^{*}$ Corresponding author
\\
$^{2}$TotalEnergies, Lacq, France}

\begin{document}
\maketitle
\begin{abstract} 
With the objective to characterize the effect of wall roughness on the flow of non-Newtonian fluids, Direct Numerical Simulations (DNS) of turbulent channel flow at a shear Reynolds number of $Re_{*}=360$ for Newtonian and Herschel-Bulkley fluids in smooth and rough channels has been performed. The rough surface was made of irregular undulations modeled with the immersed surface method. The rough surface was such that the ratio of the channel half-height to the root mean square roughness height is equal to 48, and the root mean square and the maximum crest and trough heights are equal to 7.5 and 23 wall units, respectively. This part, Part~I, is limited to Newtonian turbulent flows, serving to characterizing the roughness, and has demonstrated the generic nature of the rough surface appropriate for studying non-Newtonian fluid flows in practical applications.

The simulation results confirm that turbulence in the outer layer is not directly affected by the rough surface. The roughness effects on the turbulent stresses, the mean momentum balance and the budget of turbulence kinetic energy are confined to the layer between 0 and 25 wall units; beyond, the profiles collapse with those for smooth pipes. In the roughness  sublayer, the streamwise normal Reynolds stress is reduced while the spanwise and wall-normal components are increased. The largest increase is for the Reynolds shear stress, resulting in a significant increase in the turbulence production near the wall even though the velocity gradient is decreased. The kinetic energy budget shows that turbulence production dominates the mean viscous diffusion of turbulence kinetic energy and both mechanisms are balanced by turbulent dissipation. The friction factor using the Colebrook-White correlation calculated by specifying the sand-grain roughness as 7.5 wall units predicts the friction velocity and the bulk velocity accurately. The streaky structures that exist near smooth walls were observed to be broken by the roughness elements, leading to a denser population of coherent structures near the wall, which increase the velocity fluctuations. The coherent structures developed in the roughness layer do not seem to penetrate in to the outer layer, and no evidence could be found as to the direct impact of the roughness layer on the outer one.


\end{abstract}


\section{Introduction}
Many fluids in industrial processes are non-Newtonian and show a complex behaviour when flowing, since their resistance to flow (i.e. the ``effective viscosity'') depends on the shear stress/rate and the accumulated strain on the fluid. This non-Newtonian behaviour is due to particles or droplets, or polymers contained in the fluid, and results from the impact of the shear stress on their state of aggregation, deformation or elongation. The flow of non-Newtonian fluids is a vast and complex domain of research; this study is restricted to the flow of a non-Newtonian Herschel-Bulkley fluid in a channel with smooth and rough walls, using Direct Numerical Simulation (DNS), where the same constitutive rheology model applies over the entire domain.

The motivation for this study lies with the pressure drop over long industrial pipelines transporting Herschel-Bulkley fluids, since it determines their throughput.
From the perspective of industrial applications, the pressure drop can be computed using friction laws, which are relatively well established for non-Newtonian fluids in both laminar flow and turbulent flow in smooth pipes \citep{Metzner1955, Dodge-Metzner, Wilson1985, Wojs1993, chilton1998pressure}, even though they present some differences.
For laminar flow, the friction factor can be derived analytically; the same expression is obtained as for Newtonian fluids when the Reynolds number is replaced by a generalized Reynolds number provided by Metzner and Reed \citep{Metzner1955}. For turbulent flow, the use of a logarithmic velocity profile in the turbulent core and a laminar velocity profile in the viscous sublayer leads to the same friction factor as Prandtl's friction law for Newtonian fluids, except that it now involves the generalized Reynolds number from Metzner and Reed. Consequently, the two constants in the friction law become a function of von Karman's constant, the flow index and the thickness of the viscous sublayer, which is increased in non-Newtonian turbulent flows.

In industrial pipelines, the Reynolds number is usually moderate, and the pipe surface can be transitionally rough. The roughness is usually introduced in the friction factors for turbulent flows of shear-thinning fluids in a comparable way to the Colebrook-White friction factor for Newtonian fluids. However, experimental observations for turbulent flows of shear-thinning fluids in rough pipes do not always agree \citep{Wojs1993, Kawase1994}. Moreover, we are not aware of any work on the friction factor for shear-thinning fluids with a yield stress in rough pipes. Therefore the objective of this work is to shed light on the behaviour of Herschel-Bulkley fluids flowing in the vicinity of smooth and rough surfaces using DNS, in order to construct a physically sound wall friction correlation. For instance, we will show in this work that the friction factor for non-Newtonian turbulent flow with rough walls (with an irregular shape akin to real pipe roughness) is much higher than anticipated, due to the complex interplay between the roughness elements and the rheology.

It is perhaps important to note that it is already rare to find DNS of Newtonian turbulent flow over irregular rough surfaces. Due to the extensive nature of the work it will be described in two distinct parts. Part~I will deal with Newtonian turbulent flow over rough walls. Part~II will deal with the effect of the non-Newtonian rheology on the turbulent flow and its interaction with the rough wall.
\section{Surface roughness effects}
Historically, experiments and simulation studies of turbulent flow over rough surfaces both used idealized, easy-to-generate representations composed of sharp-edged roughness elements \citep{bailon2009turbulent,lee2011direct,leonardi2003direct,orlandi2006turbulent}. The simplification has mainly been adopted for the studies in relation to the atmospheric boundary layer. Thus, it is no surprise that in most of these studies, the roughness elements were sometimes as large as 10-15\% of the channel height, covering up to 20-30\% of the logarithmic layer! Be that as it may, abundant literature is available for turbulent channel flow with such large roughness elements – see for instance the review of \cite{jimenez2004turbulent}. DNS and Large Eddy Simulation (LES) parametric explorations could thus be conducted straightforwardly \citep{leonardi2010channel, ashrafian2004dns}, resulting in a classification of the flow behavior based on the ribs height-to-spacing ratio. Full 3D roughness elements were considered as well in some selected studies \citep{Bhaganagar,coceal2007structure,hong2011near,ma_alamé_mahesh_2021}, focusing on the effect of the Reynolds number and the spacing between elements. Since then, more irregular forms of roughness closer to what can be found in engineering systems \citep{napoli2008effect, mejia2013wall, Piomelli-roughness, van2015direct, busse2017reynolds} have been studied. Examples include degraded surfaces of heat exchangers and boilers, erosion of turbine blades, ice accumulation on aircraft wings, etc.

But rare are cases involving small and randomly dispersed roughness elements comparable in size to the inner layer, say for a wall non-dimensional unit $y^+ < 30$. In reality, the flow in the immediate vicinity of the roughness-affected layer undergoes abrupt changes, alternating ligaments breakup, stretching and elongations of the structures. This lack of attention is mainly due to the limitations of optical measurements near complex rough surfaces. In the simulation context, however, the complexity is essentially mesh-related, which has indeed motivated the resort to immersed boundary type of method. Examples include the contribution of \cite{busse2017reynolds}, who embedded surface scans of a graphite and a grit-blasted surface in the mesh. Noticeable changes in the near-wall flow have been highlighted: the viscous sublayer is shown to break down and regions of reverse flow intensify, and a ‘blanketing’ layer with mixed scaling is observed to follow the contours of the local surface, suggesting that viscous effects still pertain in this region. The destruction of the viscous sublayer is incomplete according to them. Prior to that, however, \cite{chatzikyriakou2015dns} used the immersed surface approach to cover a channel with hemispherical roughness elements of normalized roughness heights $r^+ = 10-20$. Their analysis included the effect of distribution pattern (regular square lattice vs. random pattern) of hemispherical roughness elements on the channel walls. The study revealed that the friction factor decreases with increasing Reynolds number and roughness spacing, and increases strongly with increasing roughness height. The effect of random element distribution on friction factor and mean velocities was shown to be rather weak, but a clear separation could be observed between the roughness sublayer and the outer layer, which remains relatively unaffected. 

Since the pioneering work of \cite{colebrook1937experiments}, rough-wall flow physics has been the subject of intense debate. The main consensual elements are: (i) the presence of transverse obstructions increases the drag at the wall \citep{medjnoun_rodriguez}, causing a downward shift of the velocity profile \citep{choi1993direct,orlandi2006turbulent,Chu1993ADN}; (ii) their direct effect on the velocity field decreases with increasing distance from the rough surface, and (iii) the viscous sublayer is entirely or partially destroyed \citep{busse2017reynolds}. Further, in line with classical turbulence research, most of the rough-wall flow studies have centered around the coupling between the outer flow and the flow in the near-wall region, or more precisely on the impact of the roughness on the outer layer. The challenge is to develop an inter-scale predictive model that ties the large-scale dynamics in the logarithmic region to the small-scale flow behaviour near the wall \citep{mathis2011}. According to Townsend’s wall similarity hypothesis \citep{townsend1976structure}, the flow in the outer layer, i.e. at a distance from the wall beyond a few roughness heights, is independent of the surface condition, i.e. rough-wall flow will be identical to smooth-wall flow, provided that the ratio of boundary layer thickness (or channel half-height) $\delta$ to roughness height $r$ is sufficiently large ($\delta/r > 40$ as per \cite{jimenez2004turbulent}). In support of this hypothesis, \cite{flack2005experimental,wu2007} and a number of other authors identified a merging of the mean flow, Reynolds stress and higher-order turbulence statistics in the outer layer for a range of rough surfaces. Typically, the merging is postponed to higher distances from the wall for higher-order turbulence statistics \citep{Schultz-Flack}. A merging of first- and second-order statistics has also been shown to exist for significantly low $\delta/r$ ratio of 6.75 in the context of turbulent pipe flow \citep{Chan2015} with a specific sinusoidal roughness.

The experimental campaign of \cite{squire-roughness}, addressing sandpaper roughness flows, provides sufficiently convincing elements as to the merging in the outer region of the flow, validating Townsend’s wall similarity hypothesis for the mean velocity defect and skewness profiles statistics. Outer-layer merging is also observed in the rough-wall streamwise velocity variance, but only for flows with high shear Reynolds numbers ($14'000$). Similarly, using inner variables and the roughness function to scale the flow quantities, the last-published DNS of sinusoidal rough-wall turbulence at $Re_{\tau} = 720$ of \cite{ganju2022} provided support for Townsend's hypothesis, although inner scaling fails to capture the flow physics in the near-wall region. In their contribution, \cite{wu2019} resorted to amplitude modulation analysis to explore the degree of interaction between the two layers. The analysis revealed stronger modulation effects on the near-wall small-scale fluctuations by the larger-scale structures in the outer layer, irrespective of roughness arrangement and Reynolds number. A similar conclusion has been reached by \cite{anderson2016amplitude}. A predictive inner–outer model based on exploiting principal component analysis was developed to predict the statistics of higher-order moments of all velocity fluctuations. A similar model was developed by \cite{mathis2011}.

During the course of the present study, turbulent flows of both Newtonian and non-Newtonian fluids in smooth and rough channels have been investigated using the CFD code TransAT\textcopyright. The effects of roughness and non-Newtonian rheology were first studied separately in two dedicated benchmarks, then left to act in tandem; the reference case involves Newtonian fluid in a smooth channel. In all cases, the target shear Reynolds number was fixed to $Re_{*} = 360$. Surface roughness has been created through the insertion of a CAD layer with variable roughness elements (heights and spatial distribution set as input parameters), immersed in the computational domain using TransAT’s Immersed Surfaces Technique (IST). 
 
The results will be discussed and compared in a systematic way, from smooth to rough-wall simulations for Newtonian fluids in part I, and from Newtonian to non-Newtonian fluids in Part II of the paper. The simulation campaign presented in Part I investigates the fully-developed turbulent flow over a randomly distorted wall surface, with the relative size of the roughness elements of the order of the inner layer only. The profiles of mean flow, turbulence variances, shear stresses and budgets of turbulent kinetic energies are investigated.

The simulations were carried out at TotalEnergies High Performance Computing Center on massively parallel platforms using the latest MPI standards. Typically, the simulation was distributed over 1000+ cores. The flow statistics were collected after initial transients for approximately ten flow-through times to ensure ergodic conditions were attained, in accordance with the best-practice guidelines in the area.
\section{Surface roughness characterisation}
In this study the effect of a rough surface is assessed with irregular roughness elements representative of a real pipe. The roughness was randomly generated with the average roughness height $r$ and spacing between roughness structures as parameters. The standard deviation of the roughness height $r$ was taken to be $7.5$ wall units (based on the target $Re_{*} = 360$) on each side of the mean surface plane, with the spacing between the structures as 100 wall units. 

\begin{figure}
\centering
\includegraphics[scale=0.2,angle=0]{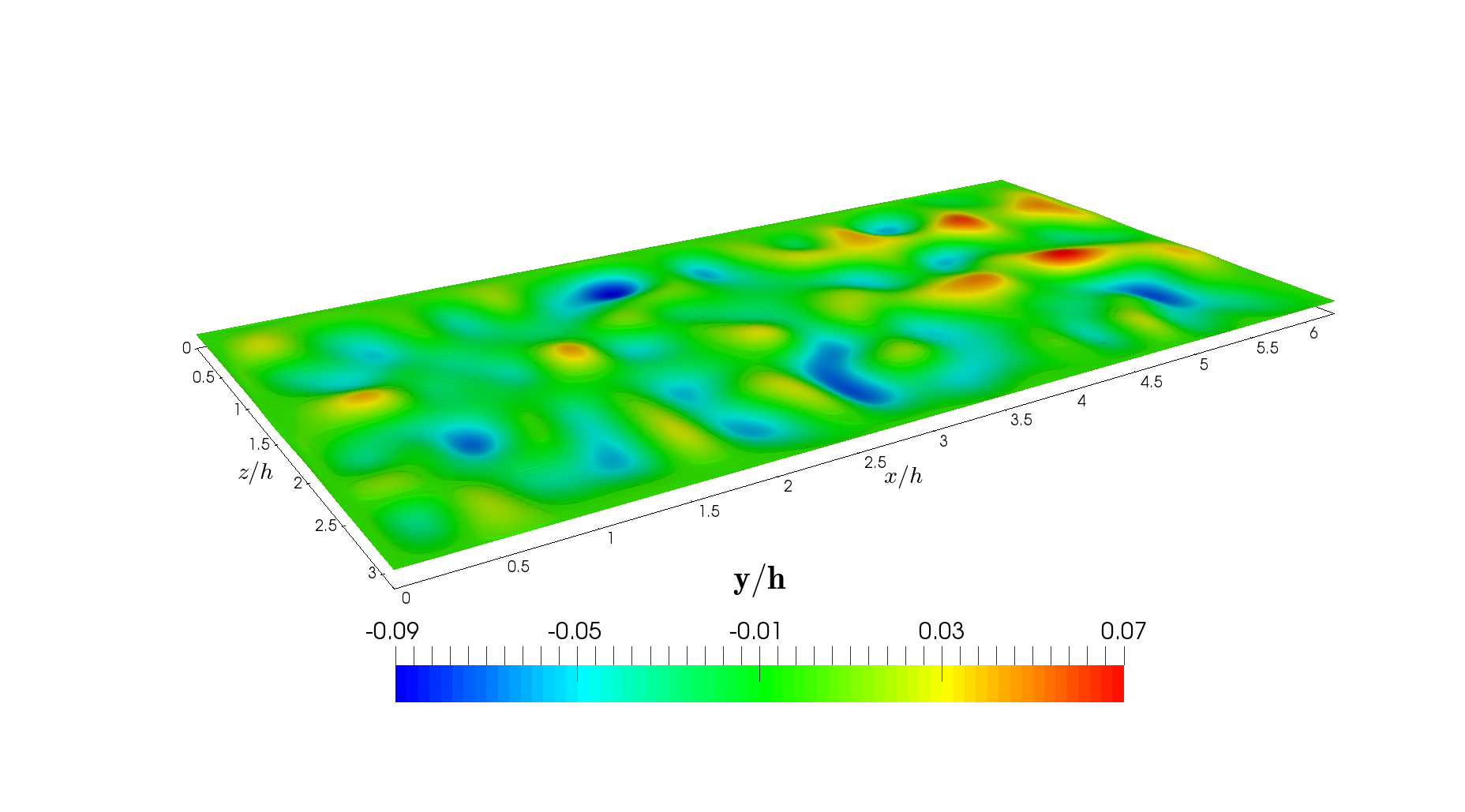}
\caption[]{Surface used for rough channel simulations}
\label{roughSurf}
\end{figure}

Using these settings a coarse roughness is generated, and then the actual rough surface is produced by smoothing the coarse surface using cubic splines interpolation. The generated surface can then be exported in a CAD format supported by the CFD software. Typically, as mentioned in the introduction, organized surface protrusions/obstructions have been considered in the literature. In this case the rough surface consists of both crests and troughs with respect to the mean wall surface at $y=0$ and $y=2\delta$ of the corresponding smooth wall simulations. One of the surfaces used in the simulations is presented in figure~\ref{roughSurf}. The maximum crest and trough were found to be $23$ wall units and the average was found to be $5$ wall units.

\begin{figure}
\centering
\includegraphics[width=0.6\textwidth]{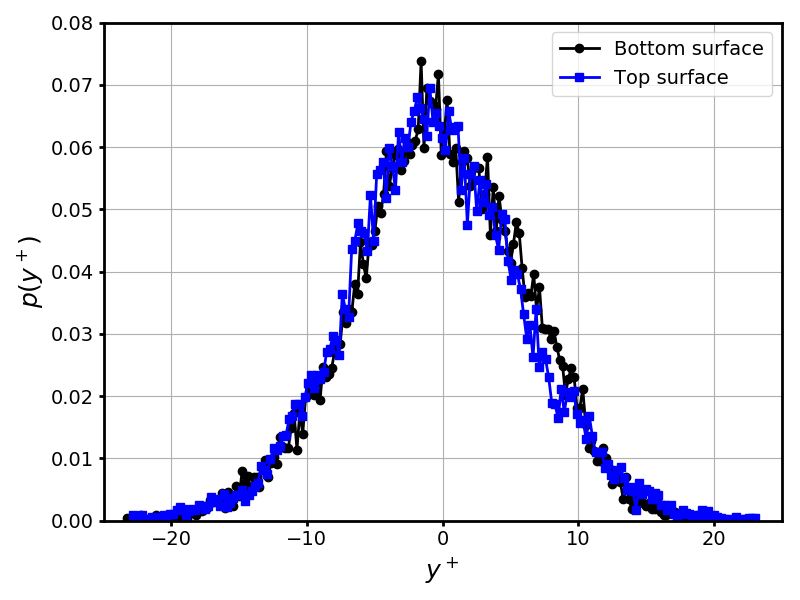}
\caption[]{Roughness distribution density function}
\label{distrib}
\end{figure}

Few references were found addressing similar roughness types, with the closest reference being the work of \cite{busse2017reynolds}, where the influence of roughness wave number filtering on the turbulent profiles is studied for two kinds of surfaces. Their roughness height distribution is similar to our work, as shown in figure~\ref{distrib} (when compared with figure~2 of \cite{busse2017reynolds}), however there are notable differences in the surface characteristics. The ratio of the channel half-height to the mean roughness is $72$ and the ratio of the channel half-height to the RMS of the roughness is $48$ in the current study compared to a highest value of  $37$ and $27$ for \cite{busse2017reynolds}, respectively. The roughness height scaled by the outer coordinate is much smaller in the present study. \cite{jimenez2004turbulent} states that universal behaviour in the outer layer should be observed for cases where $\delta/r > 40$, which is the case in this study. More specifically, this similarity hypothesis states that in the outer layer ($\delta \ge y >> \nu/u_{2};\,r$), turbulent quantities normalized by friction velocity scale are independent of the surface condition at sufficiently high Reynolds number, provided that the outer-layer thickness is much greater than the roughness height ($\delta >> r$) \citep{chung2021predicting}.

Another important difference is that their surface height distributions had significant skewness (of the two surfaces they studied, one had positive skewness and the other negative) implying that either the crests were higher than the troughs or vice versa.  In the current study the surfaces are not skewed (symmetric distribution) as shown in figure~\ref{distrib}. Additionally, it appears that our roughness has larger wavelengths (160 wall units) or lower slopes. Due to these significant quantitative differences between the surfaces, especially the ratio of the roughness to the channel half-height, the Newtonian rough wall results will be compared to \cite{chatzikyriakou2015dns} where a structured roughness was studied, but with the same solver and numerical methods. 
\section{Simulation methodology}
\subsection{Mathematical model}
The rough surface is accounted for using a variant of the Immersed
Boundary Method of \cite{peskin1977numerical} and \cite{mittal2005immersed}, known as the Immersed Surface Technique (IST).
Here, the solid object is represented in a Cartesian grid using a level set function (LSF), which we refer to as solid LSF $\phi_s(\bf x)$; positive values denote the fluid domain, negative values identify the
solid domain, and the surface of the wall is implicitly represented by $\phi_s(\bf x)=0$.
The fluid domain indicator function $H(\bf x)$ that has a value of $1$ in the fluid and $0$ within the solid varies smoothly as function of $\phi_s(\bf x)$, and the average mesh size near the fluid-solid surface $\zeta$, given as
\begin{equation}
    H(\phi_s) = \frac{1}{2} \left [ 1 + \tanh{ \left ( \frac{2 \phi_s}{\zeta} \right ) } \right ].
\end{equation}
The approach meant to resolve practical flows with complex geometries using a Cartesian mesh, can as well be used to represent individual roughness elements \citep{chatzikyriakou2015dns} or a continuous rough surface, as in the present study.

The incompressible Navier-Stokes equations are re-formulated in a strongly conservative form using the fluid-solid indicator function $H$ to yield the IST-based mass and momentum conservation equations given as
\begin{equation}
\frac{\partial }{\partial x_j} \left (H u_j \right )= 0,\\
\end{equation}
\begin{equation}
  H\frac{\partial \rho u_i}{\partial t}
  + \frac{\partial }{\partial x_j} \left ( H \rho u_i u_j \right ) = 
  - \frac{\partial p}{\partial x_i}
  + \frac{\partial }{\partial x_j} \left ( 2 H \mu S_{ij} \right )
  - 2 \mu S_{ij} \frac{\partial H}{\partial x_j},
\label{momeqn}
\end{equation}
where $\rho$ is the fluid density, $\mu$ the viscosity, $u_i$ the Cartesian velocity vector, and $S_{ij}=1/2(u_{i,j}+u_{j,i})$ the strain-rate tensor. In the IST method the no-slip condition at the immersed solid surface is achieved by rewriting the last term in equation~\ref{momeqn} as \citep{beckermann1999modeling}
\begin{equation}
- 2 \mu S_{ij} \frac{\partial H}{\partial x_j} = \mu  \frac{(u_i - u^w_i)}{\zeta} \left | 
\frac{\partial H}{\partial x_j} \right |
\end{equation}
where $u^w$ is the wall velocity which is set to zero, and $\left | {\partial H}/{\partial x_j} \right |$ is the surface area density. Note that Large-Eddy Simulations (LES) were performed on coarser grids and interpolated on to successively finer grids so as to reach the fully-developed turbulent state efficiently.

For the non-Newtonian fluid simulations, the viscosity is prescribed using the Herschel-Bulkley rheology:
\begin{equation}
    \mu = \frac{\tau_0}{\dot{\gamma}} (1 - e^{-M \dot{\gamma}}) + k \dot{\gamma}^{(n-1)},
\end{equation}
where $\dot{\gamma}$ is the magnitude of the rate of strain given as $2 \sqrt{S_{ij}\,S_{ij}}$, $\tau_0$ is the yield stress, $k$ is the consistency index and $n$ is the flow index. The above expression includes the Papanastasiou regularization \citep{mitsoulis2017numerical} to ensure that the viscosity does not attain unreasonably high values in low strain-rate regions. The constant $M$ was set to a value of $100$. However, it was noted that due to the fully-resolved turbulent fluctuations in all regions of the flow, the regularization was not applied.
\subsection{Computational algorithm}
The simulations were performed with the finite volume CFD code TransAT\textcopyright. A collocated, Cartesian grid was
used and the incompressible Navier-Stokes equations were solved. 
The mesh was locally refined in the near-wall region extending into the wall-normal direction beyond the end of the roughness and up to the buffer layer. 
The convection terms in the momentum equations are discretized using the third-order QUICK scheme \citep{leonard1995order} and the diffusion terms by second-order central differences. The higher-order convection schemes are implemented using the deferred correction approach of \cite{khosla}. The pressure correction equation is solved using the SIMPLEC pressure-correction method of \cite{van1984enhancements}. The second-order implicit Euler backward time-stepping scheme was used for the time derivative discretization given for one of the velocity components as
\begin{equation}
\left . \frac{d u}{dt}\right |^{m+1} = \frac{3 u^{m+1} - 4 u^{m} + u^{m-1}}{2 \Delta t}
\label{secondorderinTime}
\end{equation}
where $\Delta t$ is the time step. The superscript $m+1$, $m$, and $m-1$ denote the time level being solved and the two previous times, respectively. The time-step was adaptive with a Courant number of $\approx 0.3$ to guarantee good time accuracy of the simulations. For the pressure-velocity coupling the SIMPLEC algorithm was used. 
\subsection{Simulation setup}
Four DNS were performed in this study, namely, Newtonian fluid flow over a smooth wall (NS), Newtonian flow over a rough wall (NR) (both detailed in Part I), non-Newtonian flow over a smooth wall (NNS), and non-Newtonian flow over a rough wall (NNR) (discussed in Part II).
For all the simulations the domain consisted of a Cartesian box, the size of which was selected to include the largest eddies in the flow and such that the turbulent eddies would not be correlated. Thus, for the smooth wall flow the Cartesian box had dimensions $L_x = 2\pi\delta$, $L_y = 2\delta$, and $L_z =\pi\delta$, where $\delta$ is the half-channel height (wall normal direction), which was kept constant in all our simulations. Since fully developed turbulent channel flow is homogeneous in the streamwise and spanwise directions, $x$ and $z$ respectively, periodic boundary conditions were applied in these directions. No-slip boundary conditions were applied both on the upper and lower horizontal planes of the channel and on the roughness elements surface. 

The flow in the streamwise direction $x$ is driven by an constant mean pressure gradient $\Delta P = \langle -\ud p / \ud x \rangle $. The resulting shear Reynolds number is defined as $Re_{*} = {\delta u_{*}}/{\nu}$, where $u_{*} = \sqrt{ (\delta/\rho) \Delta P}$ is the shear velocity and $\nu$ is the kinematic viscosity. Variables normalized by $u_*$ and $\nu$ are denoted by the * index. For the smooth wall cases, $u_*$ is virtually the same as $u_{\tau}$, where $u_{\tau} = \sqrt{\tau_w/\rho}$ is the friction velocity, and $\tau_w$ is the average wall friction. Inner scaling is such that the variables normalized by means of $u_\tau$ and $\nu$ are denoted by the $^{+}$ index and plotted, for the Newtonian cases, against the wall unit defined as $y^+= u_{\tau} y /\nu$.

For the rough-wall cases, $u_{*}$, in addition to wall shear, also includes the effect of form drag. The mean pressure gradient was set to $-5$ for the Newtonian simulations, and to $-40$ for the non-Newtonian. Due to the non-linear rheology of non-Newtonian fluids, several iterative simulations were carried out first to estimate the mean pressure gradient that would yield a shear Reynolds number reasonably close to the target value of $360$. The final Reynolds numbers achieved for the different cases are presented in table~\ref{table:retau}. For the non-Newtonian cases the minimum spanwise-averaged viscosity has been chosen to calculate the quantities. In the case of the smooth wall, the minimum viscosity is at the wall, whereas for the rough wall, the minimum viscosity occurs slightly away from it. The $Re_*$ for the NNR case is lower than the NNS case due the higher value of the viscosity near the rough surface where the strain rate is lower than precisely at the wall for the smooth case. The wall viscosity for the NNS case was obtained to be \SI{0.24}{\pascal\second}, whereas the lowest viscosity for the NNR case was obtained as \SI{0.36}{\pascal\second}.
\begin{table}
\center
\begin{tabular}{ccccccc}
\hline
\makebox[1.1cm]{\textbf{Case}} & \makebox[1.1cm]{$u_*$} & \makebox[1.1cm]{$u_{\tau}$} & \makebox[1.1cm]{$u_b$} & \makebox[1.1cm]{$Re_*$} & \makebox[1.1cm]{$Re_{\tau}$} & \makebox[1.1cm]{$Re_b$} \\ \hline
NS  & 0.05 & 0.0500 & 0.88 & 360 & 360 & 6363 \\ \hline
NR  & 0.05 & 0.0435 & 0.70 & 360 & 313 & 5039 \\ \hline
NNS & 0.14 & 0.1400 & 2.94 & 294 & 294 & 6123 \\ \hline
NNR & 0.14 & 0.1272 & 2.33 & 197 & 177 & 3240 \\ \hline
\end{tabular}
\caption[]{$Re_{\tau}$ achieved the simulations}
\label{table:retau}
\end{table}

Turbulent statistics were computed from solution samples, once statistically ergodic conditions were obtained. Space averaging was also performed in the streamwise and spanwise directions throughout the entire domain. 
\subsection{Mesh description}
For each DNS, a sequence of Large Eddy Simulations (LES) were performed on successively refined grids for the turbulence to develop quickly on the finer grid. The coarse grid results, on reaching statistical stationarity, were interpolated on to the next finer grid. For the LES, the WALE subgrid scale model \citep{nicoud1999subgrid} was used to account for the effect of the unresolved  turbulence. For the DNS, the subgrid-scale model was switched off and simulations were carried out only with discretized Navier-Stokes equations. Meshes 1 to 4 correspond to increasingly refined LES (table~\ref{tab:smooth} for smooth wall cases), while Mesh 5 is DNS. The size of the DNS grid is consistent with a previous work \citep[][using TransAT and its IST meshing feature for roughness]{chatzikyriakou2015dns}, where it has been shown that the 15 and 26 million-cell runs return similar results in terms of first-order statistics. The advantage of IST in the particular  context of hemispherical roughness can be demonstrated by comparing the work of \cite{chatzikyriakou2015dns} to that of \cite{wu2019}, using body-fitted spectral elements (BFC). The two studies have  comparable setups in terms of roughness distribution and flow conditions. The resulting statistics point to the same observations and conclusions, although the meshes were different in size.

\begin{table}
\center
\begin{tabular}{ccccccc}
\hline
\multicolumn{2}{c}{}                                              & \textbf{Mesh 1} & \textbf{Mesh 2} & \textbf{Mesh 3} & \textbf{Mesh 4} & \textbf{Mesh 5} \\ \hline
\multirow{3}{*}{\textbf{Number of nodes}} & x  & 36 & 68 & 132 & 260 & 326 \\
                                          & y  & 18 & 34 & 66  & 130 & 163 \\
                                          & z  & 36 & 68 & 132 & 260 & 326 \\ \hline
\multirow{4}{*}{\textbf{Resolution}} & $\Delta x^+ $ & 70 & 35 & 17 & 8.8 & 4.4 \\
                                & $\Delta y_{min}^+$ & 20 & 4.3 & 1.7 & 0.2 & 0.125 \\
                                & $\Delta y_{max}^+$ & 200 & 43 & 34 & 12 & 8 \\
                                & $\Delta z^+$       & 35  & 17 & 8.8  & 4.4 & 2.2 \\ \hline
\textbf{Total}                  & ($10^6$ cells) & \textbf{0.02}  & \textbf{0.16}  & \textbf{1.15}   & \textbf{8.80}    & \textbf{17.3} \\ \hline
\end{tabular}
\newline
\caption[]{Mesh used for Newtonian and Non-Newtonian smooth walls simulations}
\label{tab:smooth}
\end{table}

The intermediate and final mesh sizes for the rough wall cases are detailed in table~\ref{tab:rough}. The mesh was refined in the undulating roughness region requiring larger number of cells in the wall-normal direction. The presence of the solid object and the IST method means that a few cells are within the solid surface as well. Note that although $u_{\tau}$ for the rough cases is lower than the smooth cases, for the sake of the normalized mesh intervals, the viscous length for $Re_{\tau} = 360$ is used.
\begin{table}
\center
\begin{tabular}{ccccccc}
\hline
\multicolumn{2}{c}{} & \textbf{Mesh 1} & \textbf{Mesh 2} & \textbf{Mesh 3} & \textbf{Mesh 4} & \textbf{Mesh 5} \\ \hline
\multirow{3}{*}{\textbf{Number of nodes}} & x  & 36 & 68 & 132 & 260 & 326 \\
                                          & y  & 25 & 47 & 77  & 152 & 191 \\
                                          & z  & 36 & 68 & 132 & 260 & 326 \\ \hline
\multirow{4}{*}{\textbf{Resolution}} & $\Delta x^+ $ & 70 & 35 & 17 & 8.8 & 4.4 \\
                                & $\Delta y_{min}^+$ & 22 & 11 & 5.4 & 2.3 & 1.7 \\
                                & $\Delta y_{max}^+$ & 195 & 41 & 26.7 & 11 & 8.6 \\
                                & $\Delta z^+$       & 35  & 17 & 8.8  & 4.4 & 2.2 \\ \hline
\textbf{Total}                  & ($10^6$ cells) & \textbf{0.03}  & \textbf{0.22}  & \textbf{1.35}   & \textbf{10.3}    & \textbf{20.3} \\ \hline
\end{tabular}
\newline
\caption[]{Mesh used for Newtonian and Non-Newtonian rough walls simulations}
\label{tab:rough}
\end{table}
\subsection{Fluid properties}
The fluid density was specified as $1000$~kg/m$^3$ for all the simulations. For the Newtonian cases the viscosity was set to $6.944 \cdot 10^{-2}$~Pa$\cdot$s.
For the non-Newtonian fluid, the parameters for the Herschel-Bulkley
law were set as $\tau_0~[\mathrm{Pa}] = 2.18 $; $k~[\mathrm{Pa}\cdot\mathrm{s}^n] = 1.42$; and $n~[-] =0.567$ which are representative of a paraffinic crude oil below the wax appearance temperature \citep{palermo2015}.
The non-Newtonian behaviour of crude oils below the wax appearance temperature can be explained by the fact that wax crystals aggregate into large porous particles, whose effective volume fraction is larger than the actual volume fraction of the wax crystals due to porosity. The size and the effective volume fraction of the aggregates is a function of the shear stress applied on them, leading to an effective viscosity dependent on the shear rate.  
\section{Validation}
\subsection{Newtonian Smooth Wall}
In order to assess the quality of the present DNS, the results were compared with other DNS reference cases of similar flows: Newtonian fluid flow in a smooth channel (labelled NS); Newtonian fluid flow in a rough channel (labelled NR); and non-Newtonian fluid flow in a smooth channel (labelled NNS). The authors could not find any reference for DNS of non-Newtonian fluid flow in a rough channel, which we'll refer to in part II as NNR. 

\begin{figure}
\centering
\includegraphics[scale=0.4,angle=0]{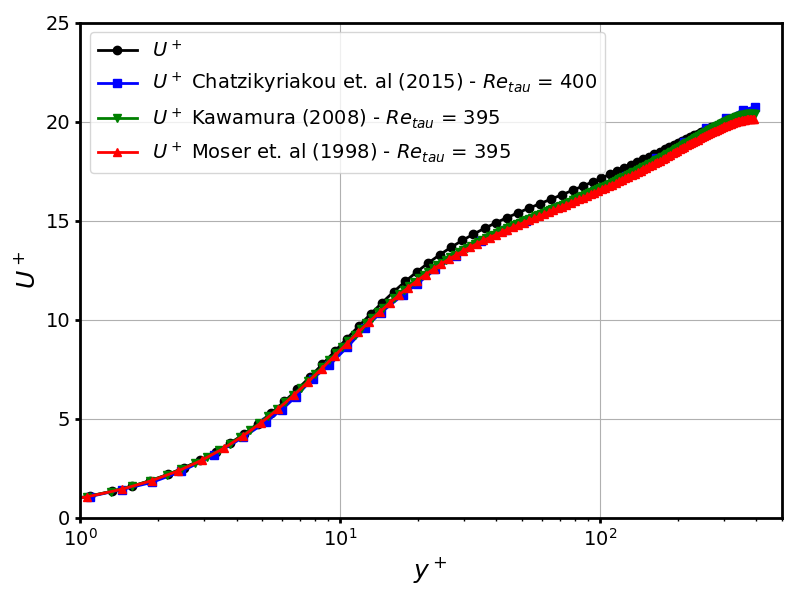}\\
\includegraphics[scale=0.325,angle=0]{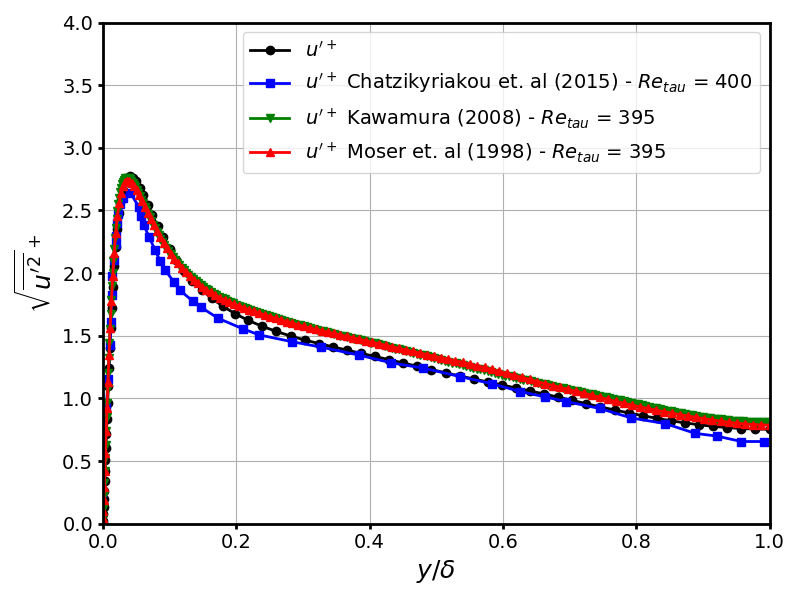}
\includegraphics[scale=0.325,angle=0]{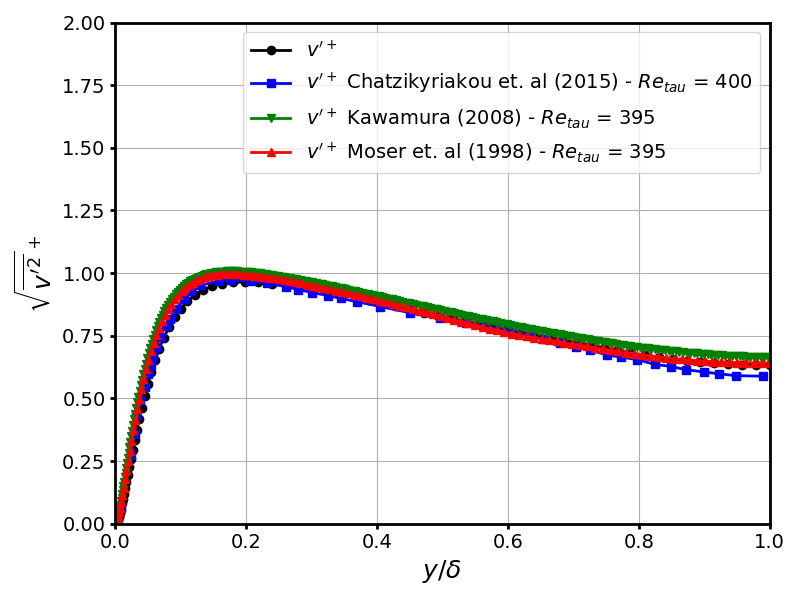}
\includegraphics[scale=0.325,angle=0]{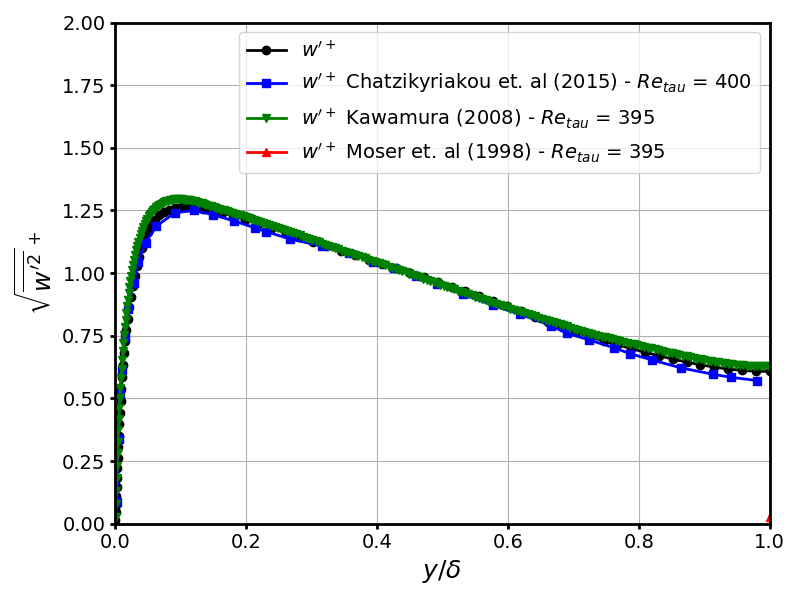}
\includegraphics[scale=0.325,angle=0]{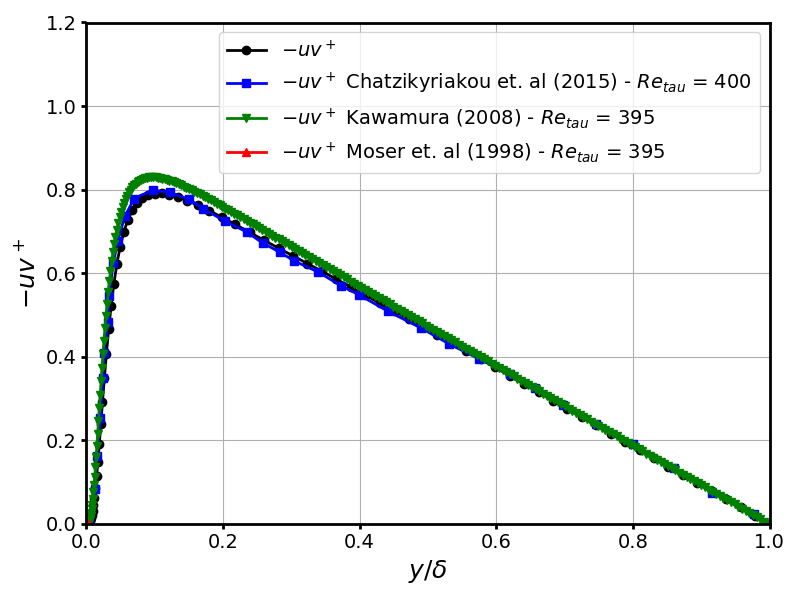}
\caption[]{Averaged profiles for a Newtonian fluid on smooth walls}
\label{results:NS}
\end{figure}

In all the reference cases, the shear Reynolds number is comparable to the present one. For NS, our results are compared to the data of \cite{kawamuraDNS} and \cite{moser_kim} at $Re_{\tau} = 395$, and \cite{chatzikyriakou2015dns} at $Re_{\tau} = 400$. As shown in figure~\ref{results:NS}, the match with reference data is good for the all the profiles. 
%
\subsection{Newtonian Rough Wall}
\begin{figure}
\centering
\includegraphics[scale=0.35,angle=0]{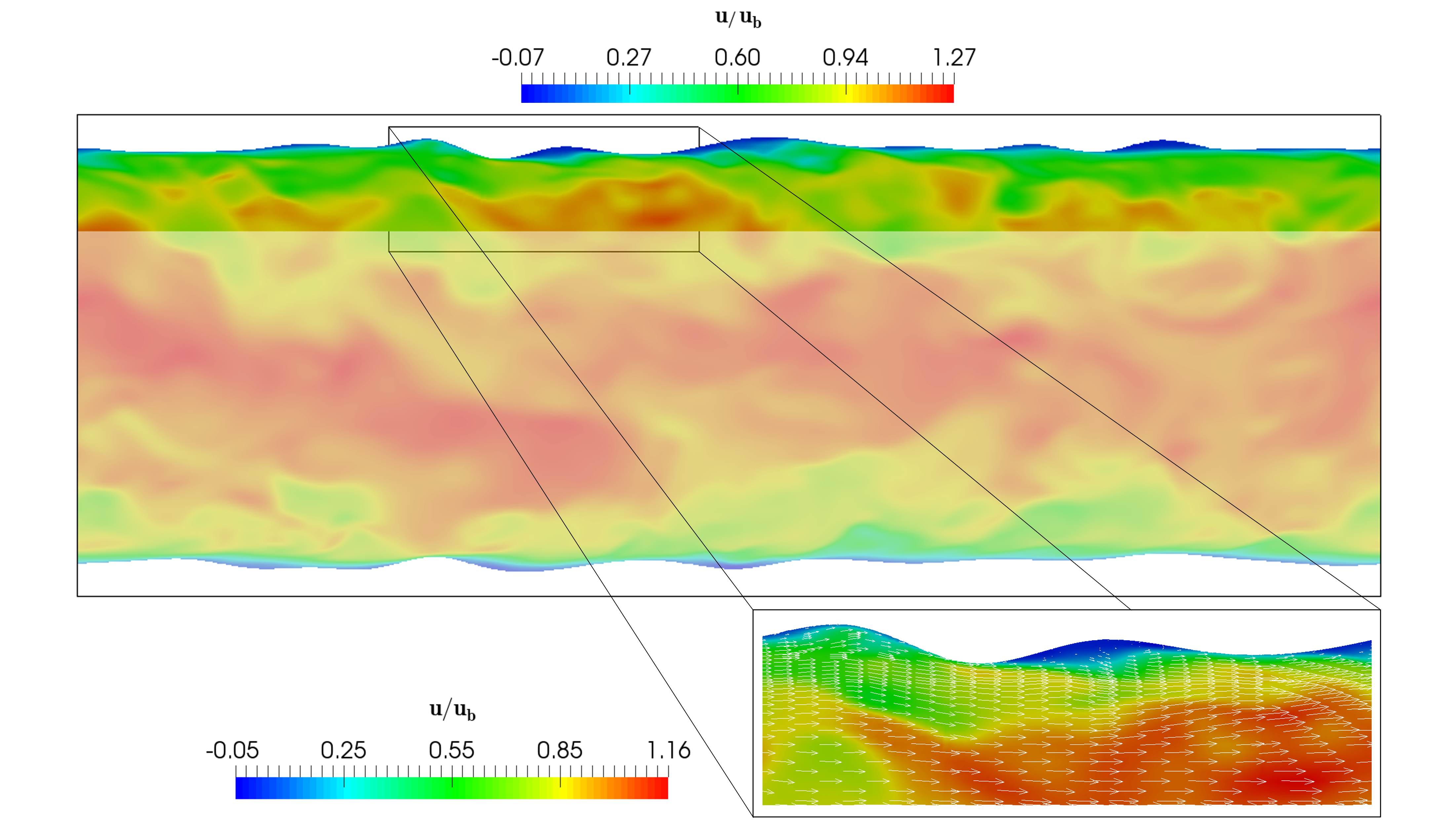}
\caption[]{Vertical slice of instantaneous velocity for NR with detail of the wall neighborhood.}
\label{fig:vertical_slice_u_NR}
\end{figure}

Prior to commenting the averaged profiles, it is perhaps useful to say a few words about the flow structure in the vicinity of the roughness layer. The instantaneous axial velocity normalized by the bulk velocity is plotted in figure~\ref{fig:vertical_slice_u_NR}. The behavior in the roughness region is different from that observed for roughness modelled as structured rows of cubic protrusions etc. The velocity shows accelerations and recirculation in the troughs, as shown in the zoomed portion of the contour plot. The roughness induces flow separation just downstream of some of the elements, which enhances form-drag contribution of the total drag.

\begin{figure}
\center
\includegraphics[scale=0.4,angle=0]{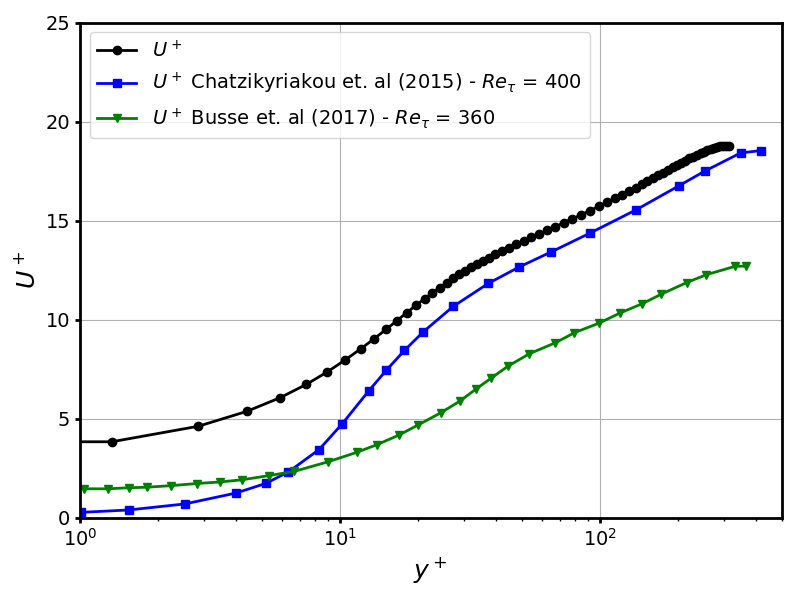}\\
\includegraphics[scale=0.325,angle=0]{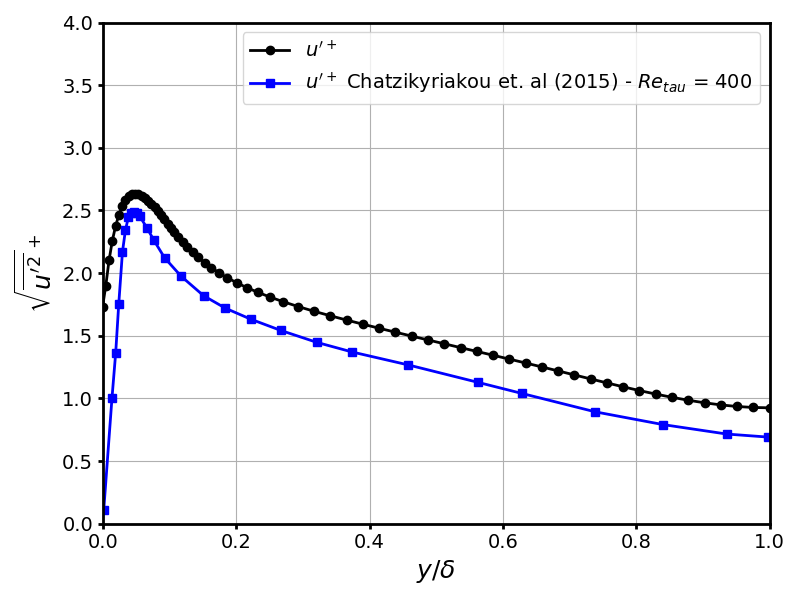}
\includegraphics[scale=0.325,angle=0]{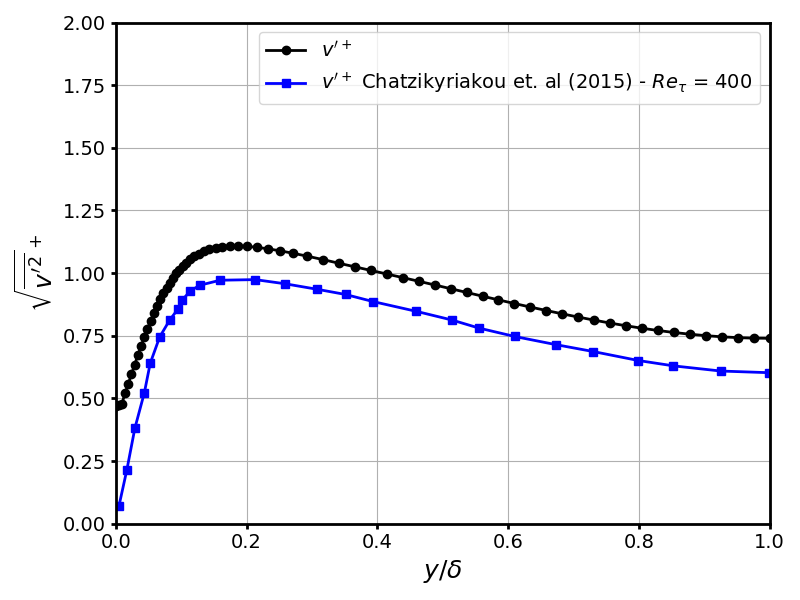}
\includegraphics[scale=0.325,angle=0]{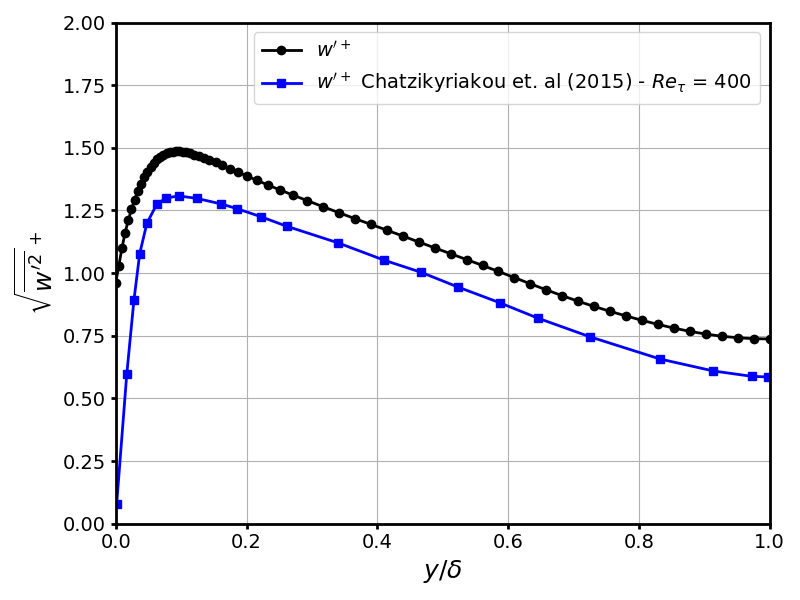}
\includegraphics[scale=0.325,angle=0]{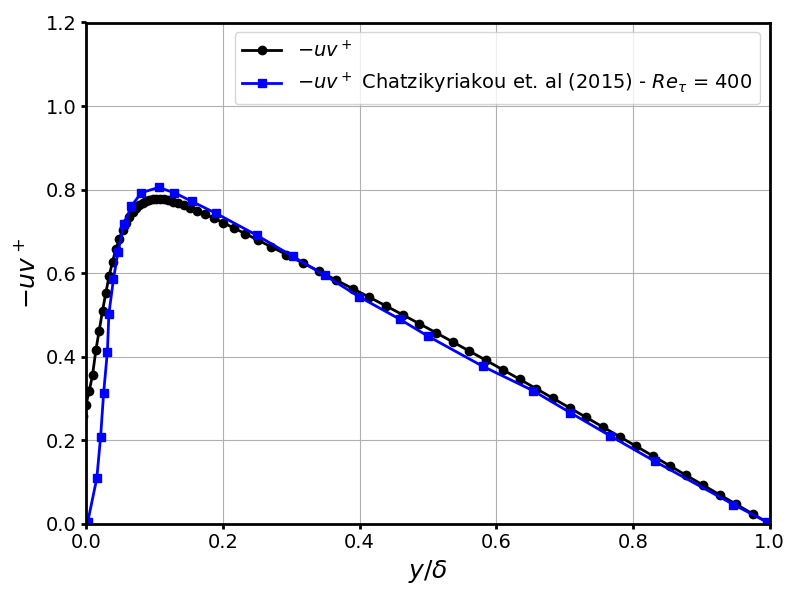}
\caption[]{Profiles for a Newtonian fluid on rough walls}
\label{results:NR}
\end{figure}

The mean velocity, turbulent fluctuations and turbulent stress profiles for the rough wall case are compared to the DNS data of \cite{chatzikyriakou2015dns} and \cite{busse2017reynolds} in figure~\ref{results:NR}. Note that coincidentally, the two references report using the same immersed boundary/surface technology for meshing the roughness elements.

As to the mean velocity profile, significant differences with the results of \cite{busse2017reynolds} can be observed. Namely, the effect of roughness elements is very strong in their case due to the much higher and steeper characteristics of the roughness used. The outer behaviour obtained in the present study matches with the results of \cite{chatzikyriakou2015dns} for hemispherical protrusions, as expected, based on the $\delta/r$ criterion discussed earlier. The viscous layer has higher flow in the current study due to the fact that the troughs go below the $y=0$ line, whereas in the case of organized hemispherical protrusions, the velocity has to be $0$ at $y=0$. Therefore, in this case we primarily verify the outer layer behaviour by comparing to the mean velocity profile of \cite{chatzikyriakou2015dns}.

The turbulent statistics show a good match in the outer region, although a higher overall level of turbulence is observed in the current study; due to the higher roughness heights (RMS of 7.5 and maximum heights of 23 wall units). In the inner region, especially at $y=0$ much higher levels are observed in the current study due to the fact that the troughs extend further down (as low as 23 wall units), therefore the mean velocity need not be zero at the average wall plane.  
%
\section{Newtonian Fluid Results}
\subsection{Turbulence statistics}
%
%
The results presented in this work are space-averaged (in the streamwise and spanwise directions) and time-averaged until no further evolution in time is observed.
\subsubsection{Mean velocity profile}
\begin{figure}
\centering
\includegraphics[scale=0.4,angle=0]{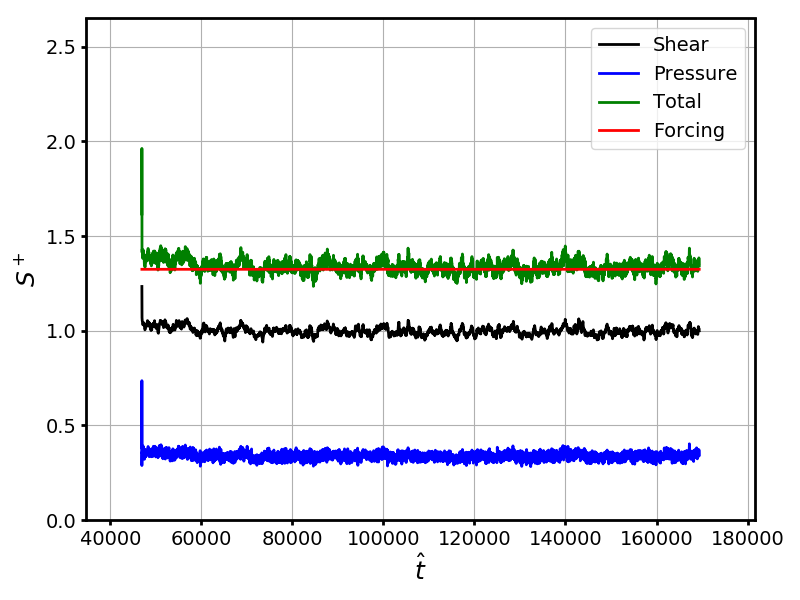}
\caption[]{Time variation of normalized viscous drag and form drag.}
\label{dragbalance}
\end{figure}

The roughness causes an increase in total drag (viscous and form drag contributions) resulting in a lower $Re_{\tau}$ for the same pressure forcing which also results in a lower bulk (and average centerline) velocity. In fact, in the present case, the mean velocity goes slightly negative in the trough regions as shown in the zoom of figure~\ref{fig:vertical_slice_u_NR}. It was observed that the viscous shear at the wall accounts for 75\% of the applied pressure forcing and 25\% is accounted for by form drag; this dynamic force balance is presented in figure~\ref{dragbalance}. The reduction in the bulk and shear velocities results in a lower shear Reynolds number of $313$.

\begin{figure}
\centering
\includegraphics[scale=0.4,angle=0]{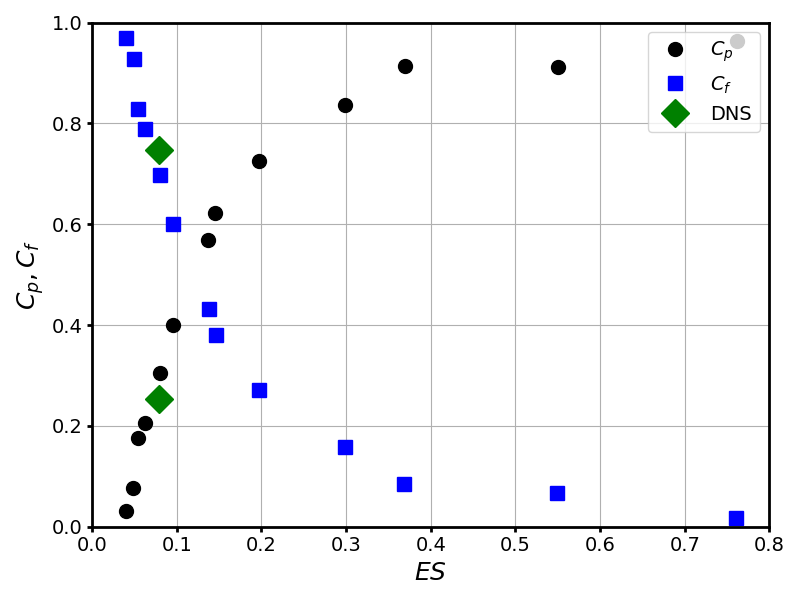}
\caption[]{Comparison of the form drag coefficient versus the effective slope with data from \cite{napoli2008effect}.}
\label{effectiveslope}
\end{figure}

One of the increasingly validated hypothesis in rough-wall turbulence is the proposal by \cite{napoli2008effect} that relates the ratio between form drag and total drag to the {\em effective slope} of the rough surface. \cite{van2015direct} calculated the form drag to total drag ratio versus effective slope with different roughness topologies, and obtained the same shape as \cite{napoli2008effect} further reaffirming the hypothesis that this could be a {\em universal} behaviour. The effective slope is defined as
\begin{equation}
    ES = \frac{1}{L_x} \int_0^{L_x} \left | \frac{\partial r}{\partial x} \right | dx
\end{equation}
where the integration is along the streamwise direction and $L_x$ is the length of the domain. The effective slope in the present case was found to be $0.083$ in the streamwise direction and $0.075$ along the spanwise direction. The data from \cite{napoli2008effect} along with the data obtained from the present study are shown in figure~\ref{effectiveslope}. The present data falls very much within the data of \cite{napoli2008effect} thereby reaffirming the hypothesis of a universal behaviour of rough walls. 
 
The roughness-induced form-drag contribution to the overall drag, affects the mean velocity profile such that the log law can be rewritten as \citep{durbin2001rough}
%
\begin{equation}
U^+(y^+) = \frac{1}{\kappa}  \ln(y^+) + B + \Delta B_r (r^+) 
\end{equation}
where $\kappa$ is the von Karman constant and $B$ is the intercept for smooth walls. The shift in the mean velocity profile due to roughness is denoted by $\Delta B_r$, and depends on the roughness height $r^+$. The function $\Delta B_r$ for intermediate roughness $2.25 \leq r^+ \leq 90$ is given as
\begin{eqnarray}
    \Delta B_r(r^+) &=& \xi(r^+) \left [ 8.5 - B - \ln(r^+)/\kappa\right ], \nonumber \\
    \xi(r^+) &=& \sin \left (\frac{\pi/2 \ln(r^+/2.25)}{\ln(90/2.25)} \right ).
    \label{roughlawofwall}
\end{eqnarray}

\begin{figure}
\center
\includegraphics[scale=0.4,angle=0]{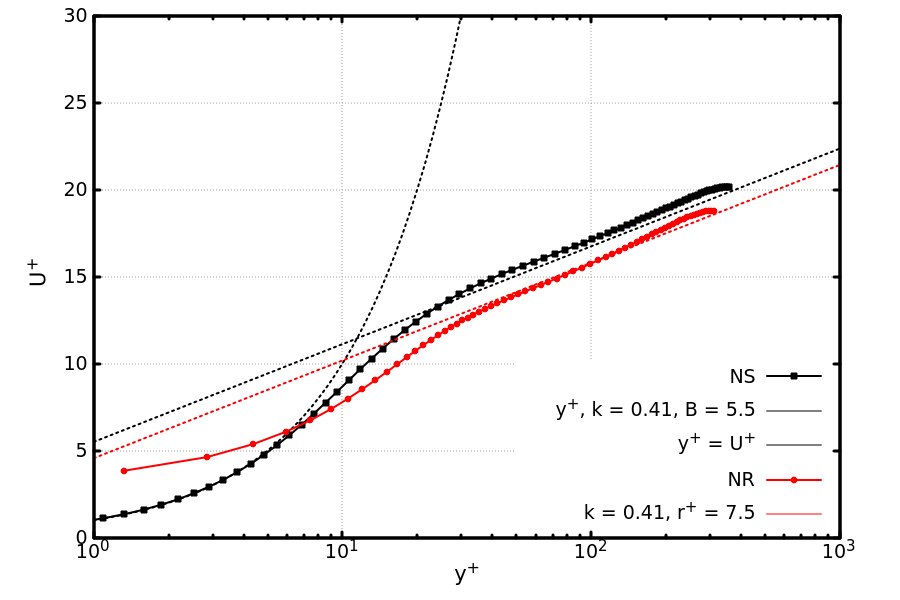}
\caption[]{Mean velocity profile: smooth vs. rough wall.}
\label{results:newtoniancomparison}
\end{figure}

Figure \ref{results:newtoniancomparison} reports the mean profiles, including the smooth- and rough wall cases at $Re_{\tau} = 360$, and $313$, respectively, in wall coordinates. As can be observed, a very good match to the modified log law is obtained by using smooth-wall values of $\kappa = 0.41$, $B = 5.5$, and $r^+ = 7.5$ in equation~\ref{roughlawofwall}. Others \citep{endrikat2022reorganisation} fit the mean velocity profile by decreasing the von Karman constant $\kappa$ to a value of $0.3$ for the case of riblets with a height of $60$ wall units. A good match is obtained in this case using a generic roughness adjustment without the need to adjust $\kappa$ shows that the roughness characteristics chosen in this study are of a generic nature and therefore appropriate for the study of roughness effects on non-Newtonian flows in pipelines.

\begin{figure}
\centering
\includegraphics[width=0.49\textwidth]{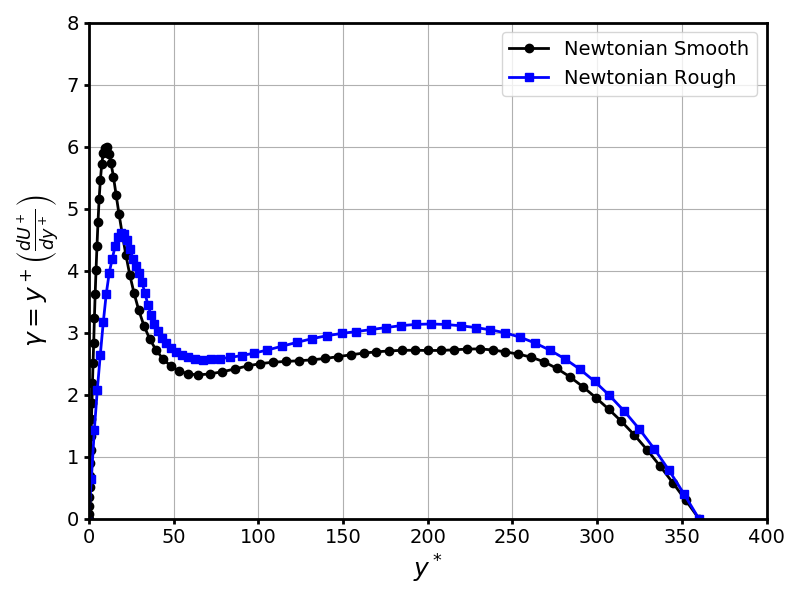}
\includegraphics[width=0.49\textwidth]{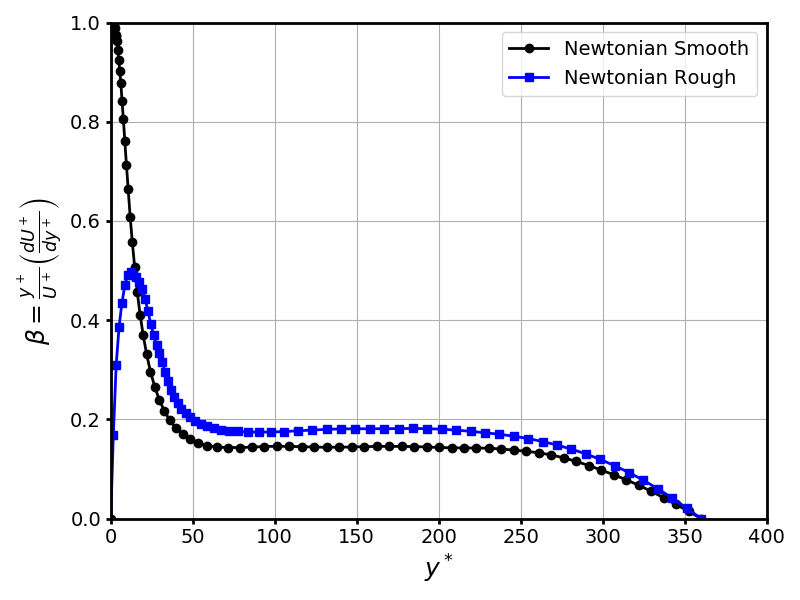}
\caption[]{Structure of boundary layer, (left) Diagnostic for log law (right) Diagnostic for power law $\beta$}
\label{fig:beta-gamma-NS-NR}
\end{figure}

The structure of the turbulent boundary layer with regard to satisfying a logarithmic law or a power law can be analyzed by plotting these two profile-characteristics parameters $\gamma = y^+ \partial U^+/\partial y^+$, and $\beta = (y^+/U^+) \partial U^+/\partial y^+$, respectively. Figure~\ref{fig:beta-gamma-NS-NR} shows that for the rough surface, the interval where a log-law behaviour can be expected is smaller with a lower effective $\kappa$. On the other hand, the variation of $\beta$ shows that for both the smooth and rough cases, the mean velocity profile could be very well represented as a power law. This behaviour is similar the grit-blasted roughness presented by \cite{busse2017reynolds}.
\subsubsection{Turbulent stress profiles}
\begin{figure}
\center
\includegraphics[scale=0.325,angle=0]{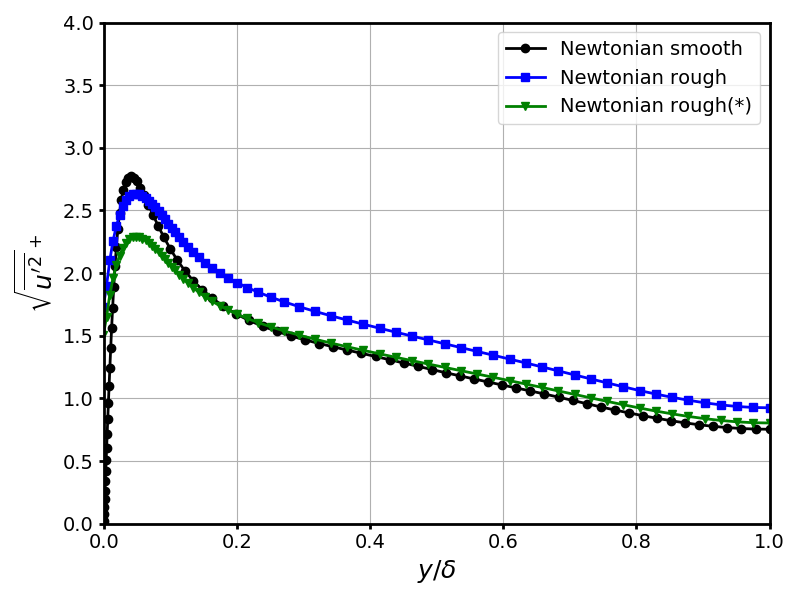}
\includegraphics[scale=0.325,angle=0]{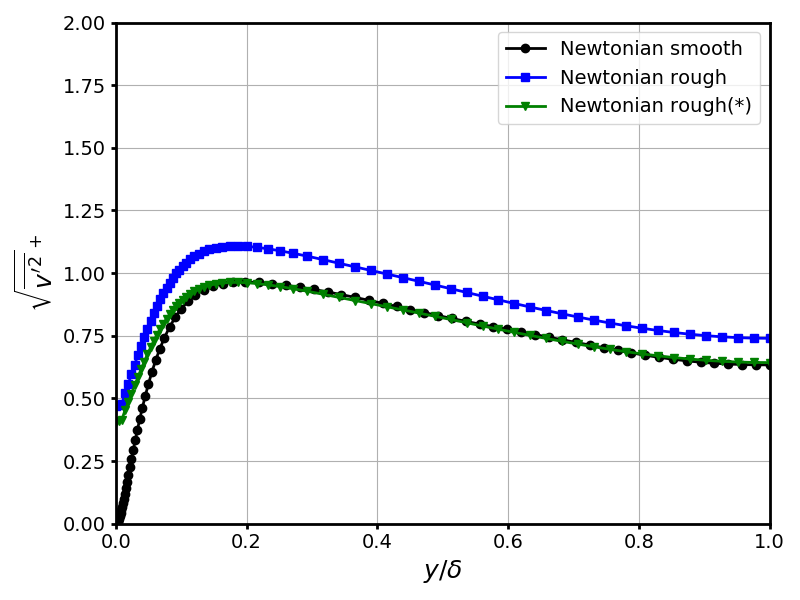}
\includegraphics[scale=0.325,angle=0]{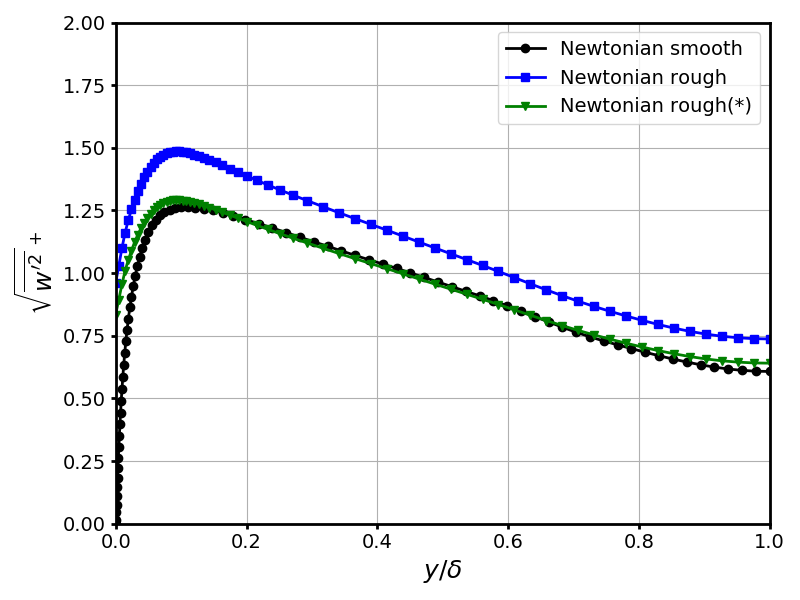}
\includegraphics[scale=0.325,angle=0]{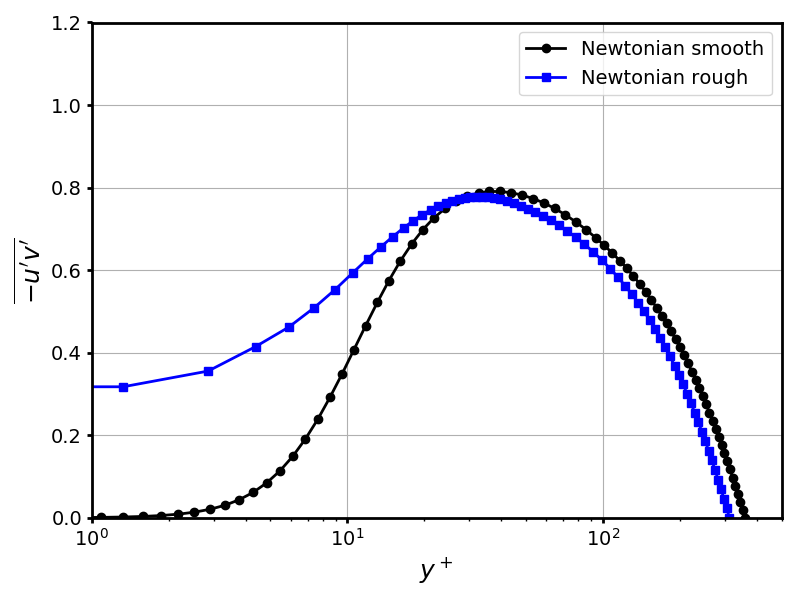}
\caption[]{RMS of velocity with outer scaling and turbulent stress with inner scaling: smooth vs. rough wall results}
\label{results:newtoniancomparison_outer}
\end{figure}

We present the normal stresses normalized by both $u_{\tau}$ and $u_*$ in figure~\ref{results:newtoniancomparison_outer}. When normalized by the friction velocity $u_{\tau}$, the normal turbulent stresses with rough walls are consistently higher than the smooth wall case over the entire domain, except for the streamwise component close to the wall where the peak value is slightly lower. However, when normalized by $u_*$, in the region $0 < y^+ < 25$, the streamwise component peak is significantly lower than the smooth wall, whereas the cross-stream components are higher. Away from the inner layer, the normal stresses match the smooth wall almost exactly. This shows that for the roughness height studied, the effect of the roughness is not felt in the outer region. Therefore, for a given driving force (pressure gradient), increased turbulence production and intensity is restricted to the inner region, consistent with the observations of \cite{Bhaganagar} and \cite{ma_alamé_mahesh_2021}. 

The variation of the turbulent shear stress with inner scaling is presented in figure~\ref{results:newtoniancomparison_outer}. The marked enhancement in the turbulent shear stress in the near-wall region results in an increased turbulence production even though the mean velocity gradient is lower as compared to the smooth wall. The net result, however, is an increased turbulence production.

\cite{flack2005experimental} indicate that $\delta/r_s$ (where $r_s$ is the equivalent sand-grain roughness height) and not $\delta/r$ is the proper parameter to indicate if the roughness effect on the turbulence will be strong or weak. The comparison to the Colebrooke-White correlation (section~\ref{wallfriction}) suggests that in the current study the sand-grain roughness is almost the same as the roughness height, essentially showing that the chosen rough surface represents a generic roughness. As already mentioned earlier, the $\delta/r$ in the current study is approximately $70$.
\subsubsection{Energy Spectra}
\begin{figure}
\centering
\includegraphics[width=0.49\textwidth]{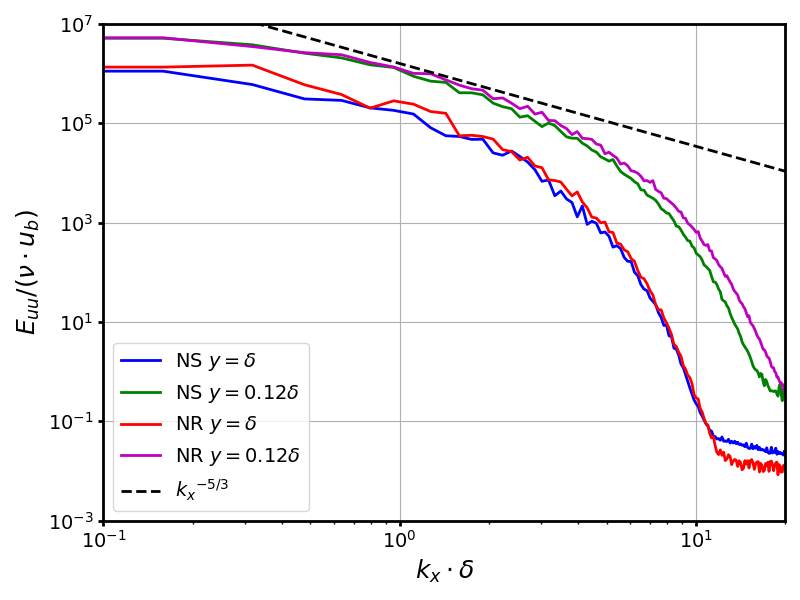}
\includegraphics[width=0.49\textwidth]{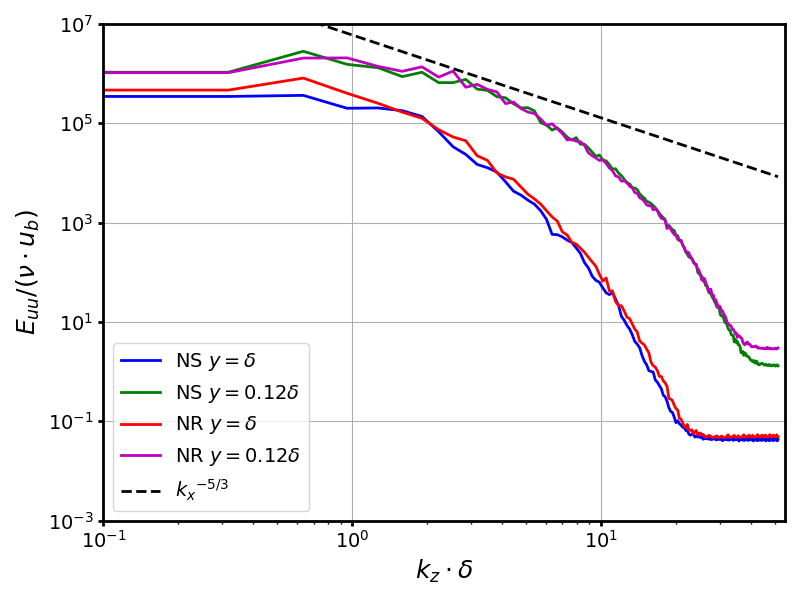}
\includegraphics[width=0.49\textwidth]{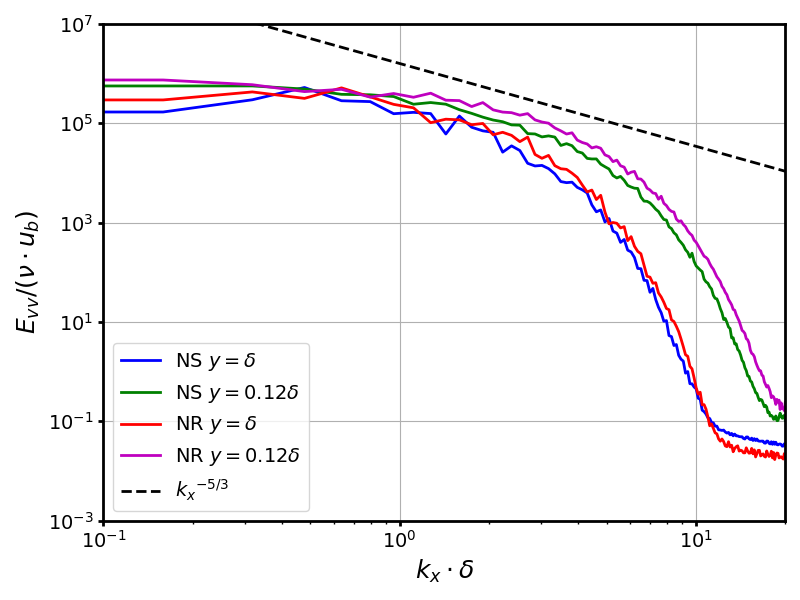}
\includegraphics[width=0.49\textwidth]{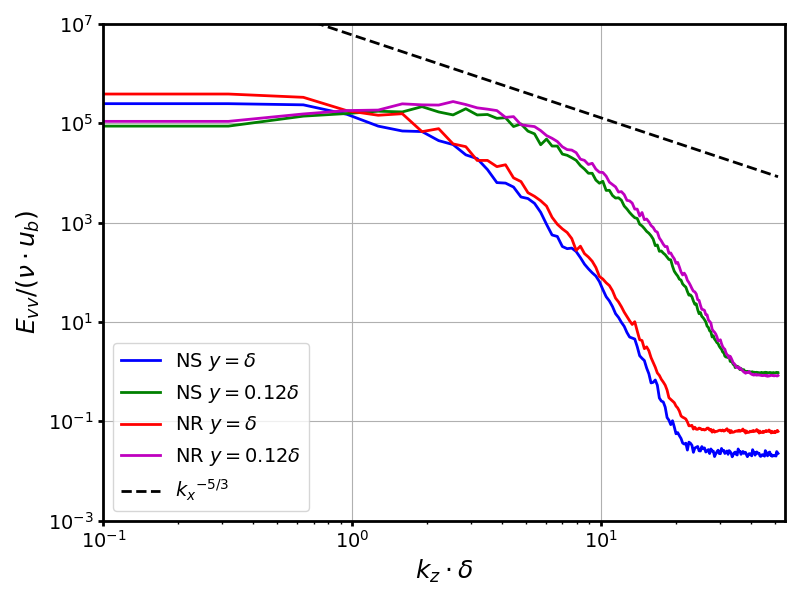}
\caption[]{Left: PSD in the x-direction for NS case compared to NR case. Right: PSD in the z-direction for NS case compared to NR case at $y/\delta = 0.12$, and $1$.}
\label{fig:energy_spectra}
\end{figure}

The power spectral density (PSD) of the streamwise and the wall-normal velocities in the streamwise and spanwise directions at two locations ($y/\delta = 0.12$, and $1$) are presented in figure~\ref{fig:energy_spectra}. The locations chosen based on the work of \cite{mitishita2021} to which the results for the non-Newtonian cases will be compared. The spectra of the spanwise velocity is very similar to the wall-normal velocity and are therefore not presented here. Firstly, we note that the overall energy near the wall is higher than at the center of the channel. The $-5/3$ slope is observable for a short wavenumber band in the mid-range of wavenumbers. The spectra also show that the simulations are well resolved with a steep drop in the energy in the dissipative high wavenumber range. At $y/\delta = 1$ (centerline) no impact of the roughness is detected over the whole resolved wavenumber range. 

Near the wall ($y/\delta = 0.12$) a consistent increase in the energy at higher wavenumbers is observed in the streamwise direction indicating reduction in streamwise coherence of the streamwise fluctuations; on the other hand no impact of roughness is observed for the spanwise spectra of the streamwise fluctuations. Even though the rough surface has undulations both in the streamwise and spanwise directions, it does not have any impact in the spanwise coherence of the streamwise velocity. The same observations can be made for the wall-normal velocity component, showing an increase in energies at the higher wavenumbers near the wall in the streamwise spectra. However, a slight increase is also observed in the spanwise spectra.

\subsubsection{Mean stress balance}
The mean stress balance shows the relationship between the mean viscous shear stress and the turbulent shear stress. For a non-Newtonian fluid, an additional term due to fluctuations in the viscosity arises \citep{singh2017influence} (to be discussed in Part II). For a smooth surface, the viscous shear is maximum at the wall and the turbulent shear stress is zero. At the channel center line, both the stresses are zero due to symmetry. 

\begin{figure}
\centering
\includegraphics[scale=0.325,angle=0]{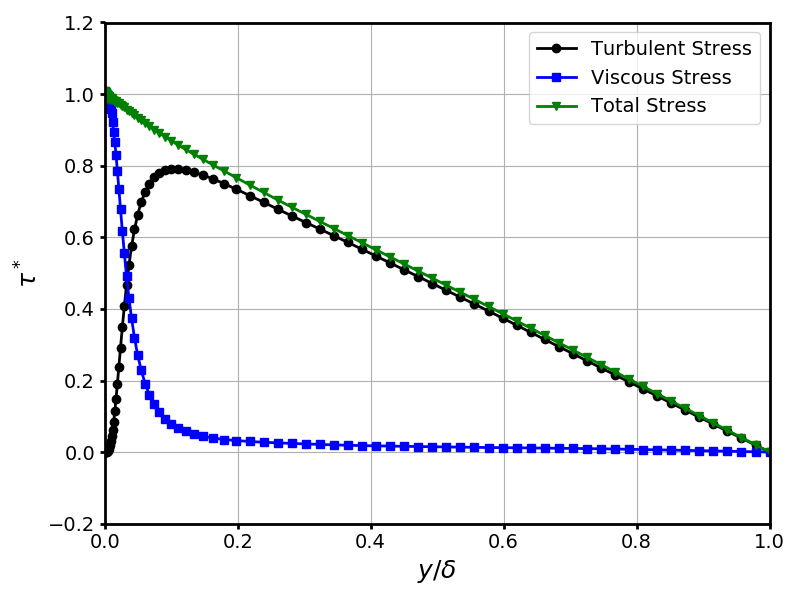}
\includegraphics[scale=0.325,angle=0]{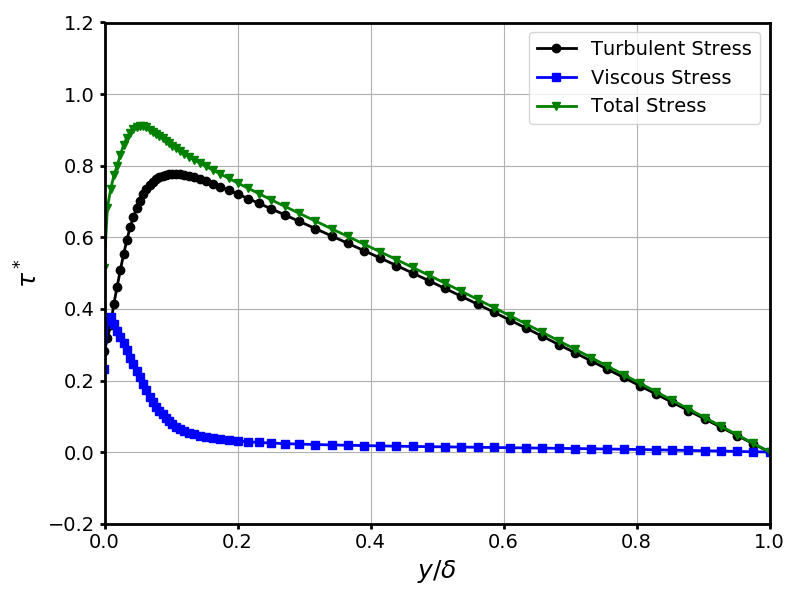}
\caption[]{Mean shear stress balances for Newtonian fluid with outer scaling: smooth wall vs. rough wall.}
\label{shearStressBalance-NS-NR}
\end{figure}

\begin{figure}
\centering
\includegraphics[scale=0.325,angle=0]{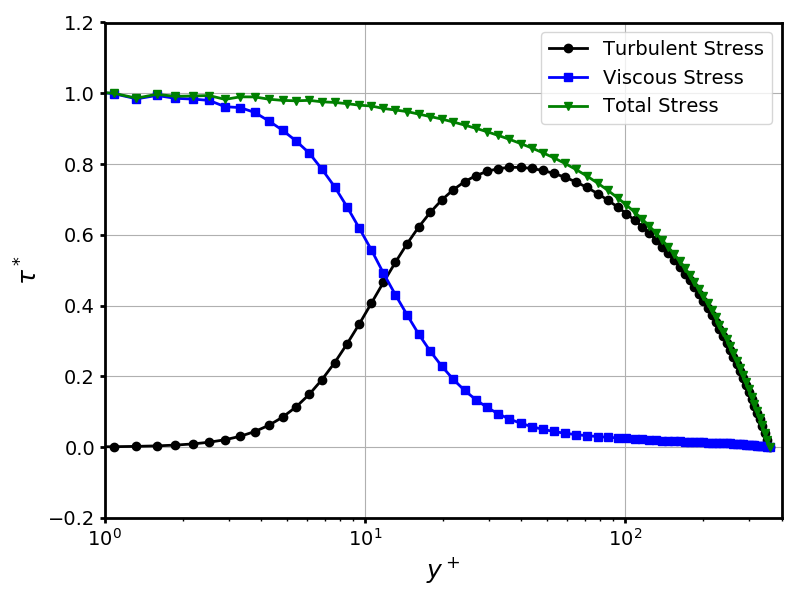}
\includegraphics[scale=0.325,angle=0]{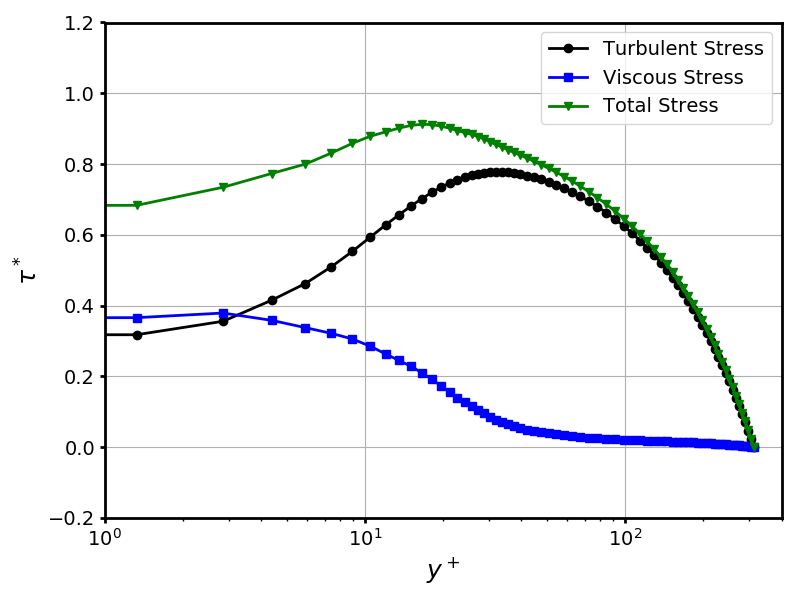}
\caption[]{Mean shear stress balances for Newtonian fluid with inner scaling: smooth wall vs. rough wall.}
\label{shearStressBalanceInner}
\end{figure}

The mean shear stress balances for the Newtonian cases are presented in figure~\ref{shearStressBalance-NS-NR}. The same plot is shown in log scale in figure~\ref{shearStressBalanceInner} in order to zoom in to the near-wall region. Note that, for the rough wall case, the terms have to be normalized by $u^*$ in order to compare to the smooth wall case. For the NS case, the stress balance is as expected and can be considered as a validation of the results. For the rough channel, the total stress peaks away from $y=0$; at the wall the viscous shear is of the same magnitude as the turbulent stress. The point at which the two stress components are in balance for a smooth wall gets shifted towards the wall. The shift that is approximately $7.5$ wall units matches the RMS of the wall roughness, however in the absence of simulations with different roughness heights, this could be a mere coincidence. In the outer scaling it appears that the effect of roughness is not visible beyond $y/\delta = 0.1$.
\subsubsection{Mean turbulent kinetic energy budget}
The mean kinetic energy equation for general non-Newtonian fluids is presented in equation~\ref{tkebudget}. The terms in equation~\ref{tkebudget} are annotated as follows; turbulence production $P^+$, turbulent transport $T^+$, pressure diffusion $\Pi^+$, mean viscous diffusion $D^+$, and mean viscous dissipation $e^+$. In these quantities, $S_{ij}$ and $s_{ij}$ denote the mean and fluctuating rate-of-strain tensors, respectively. The four additional terms resulting from the non-Newtonian rheology have been annotated as in \cite{singh2017influence}, and are named as follows, where the subscript $vv$ denotes {\em variable viscosity}. $\xi_{vv}$: mean shear turbulent viscous transport, $\chi_{vv}$: mean shear turbulent viscous dissipation, $\mathcal{D}_{vv}$: viscous turbulent transport, and $\epsilon_{vv}$: turbulent viscous dissipation. The additional variable viscosity terms and the effect of roughness on those terms will be discussed in Part II.

\begin{eqnarray}
\frac{\partial k}{\partial t}
+ \bar{u}_j\frac{\partial k}{\partial x_j} &=&
-\underbrace{\overline{u'_i u'_j} \frac{\partial \bar{u}_i}{\partial x_j}}_{\mathrm{Production}}
-\underbrace{\frac{\partial \overline{u'_i u'_i u'_j}}{\partial x_j} }_{\mathrm{Turb.\,Transport}}
- \underbrace{\frac{1}{\rho} \frac{\partial \overline{p'u'_i}}{\partial x_i}}_{\mathrm{Pressure\,Diff.}} \nonumber \\
&&+ 
\underbrace{\frac{\partial }{\partial x_j}  \left (2 \overline{\nu}\, \overline{u'_i s_{ij}} \right )}_{\mathrm{Mean\,Visc.\,Diff.}}
- \underbrace{2 \overline{\nu} \,\overline{s_{ij} \frac{\partial u'_i}{\partial x_j}}}_{\mathrm{Mean\,Visc.\,Diss.}} \nonumber \\
&&+ 
\underbrace{\frac{\partial }{\partial x_j}  \left (2 \overline{u'_i \nu'} \,\,\overline{S}_{ij} \right )}_{\xi_{vv}}
- \underbrace{2  \overline{\nu' \frac{\partial u'_i}{\partial x_j}} \,\,\overline{S}_{ij}}_{\chi_{vv}} \nonumber \\
&&+ 
\underbrace{\frac{\partial }{\partial x_j}  \left (2\, \overline{\nu' u'_i s_{ij}} \right )}_{\mathcal{D}_{vv}}
- \underbrace{2\,  \overline{\nu' s_{ij} \frac{\partial u'_i}{\partial x_j}}}_{\epsilon_{vv}}
\label{tkebudget}
\end{eqnarray}

\begin{figure}
\centering
\includegraphics[width=0.49\textwidth]{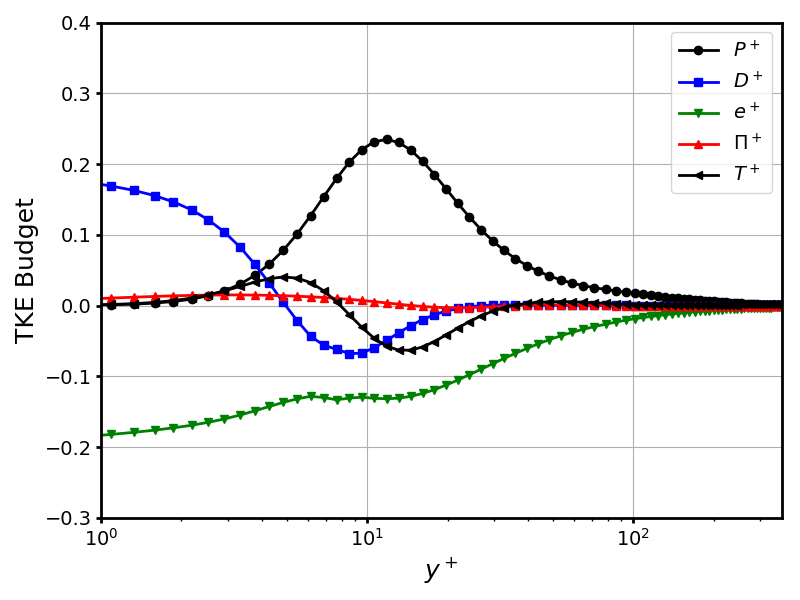}
\includegraphics[width=0.49\textwidth]{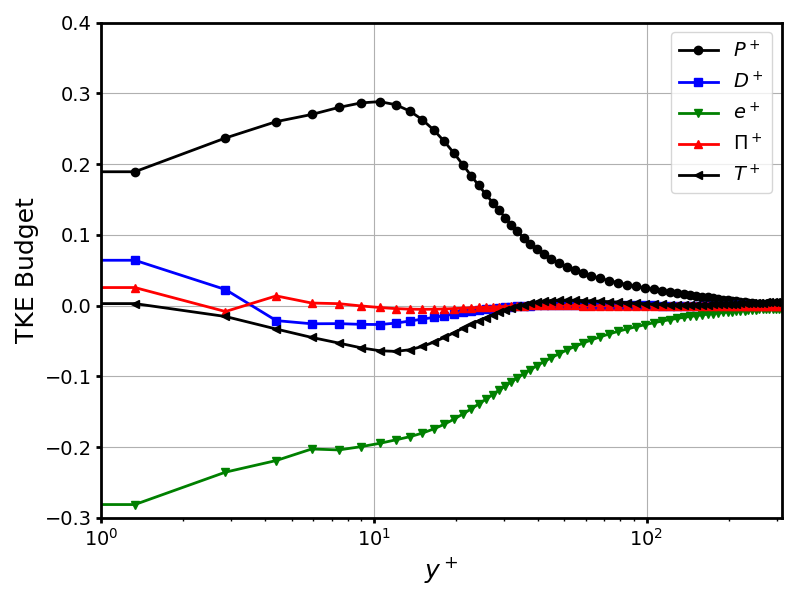}
\includegraphics[width=0.49\textwidth]{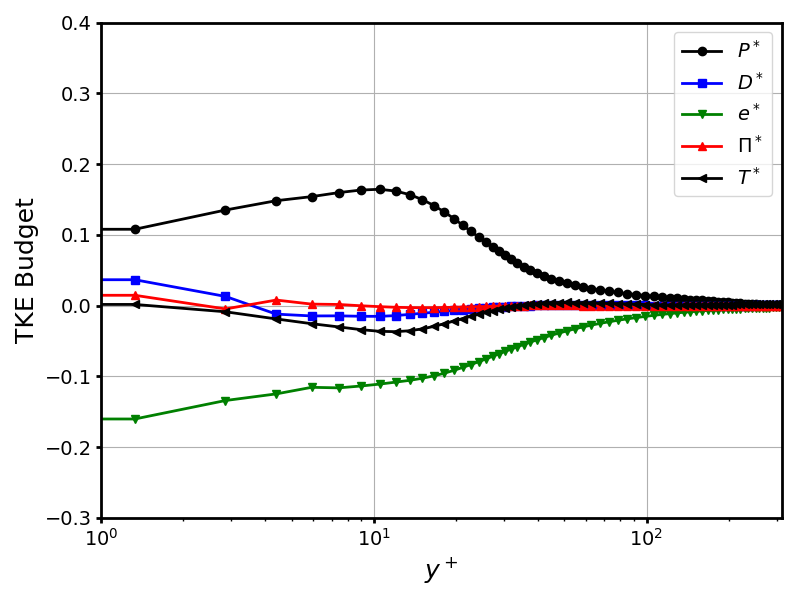}
\includegraphics[width=0.49\textwidth]{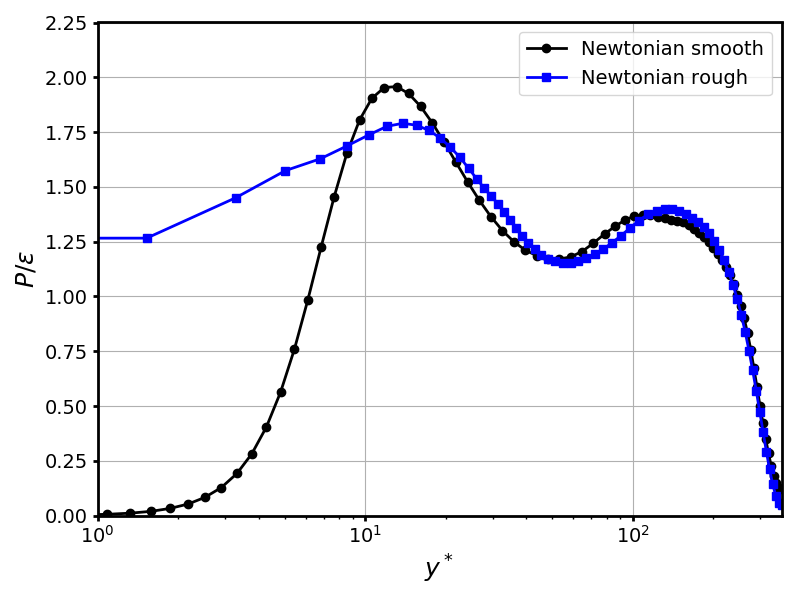}
\caption[]{Turbulent kinetic energy balance for the Newtonian smooth case (top left) and for the Newtonian rough case in + units (top right) and in * units (bottom). Comparison of turbulent shear rate $P/\varepsilon$ between smooth and rough walls.}
\label{fig:stat20}
\end{figure}

The turbulent kinetic energy (TKE) budgets for the smooth and rough wall cases are presented in figure~\ref{fig:stat20}(a) and (b), respectively. The main observation is that the production of TKE is significantly higher in the viscous layer for rough walls. For smooth walls, the main balance in the viscous sublayer is between mean viscous diffusion and dissipation with turbulence production going to zero rapidly as the wall is approached. For rough walls, turbulent dissipation balances the sum of turbulence production and mean viscous diffusion (to a smaller extent).
Also note that turbulence production is much larger than the mean viscous diffusion for rough walls. The pressure diffusion and turbulent transport terms are smaller in magnitude in both cases. 

Note that, when the TKE budget is plotted in terms of $^{+}$ units, the magnitudes of the terms are larger for the rough wall case due to the smaller normalization factor given as $u_{\tau}^4/\nu$ with a lower value for $u_{\tau}$. When the TKE budget is plotted in terms of * units, it can be seen that the differences between rough and smooth walls are negligible for $y^+ > 25$, which supports Townsend's hypothesis that the turbulence in the outer layer is unaffected by the inner layer. Moreover the profiles of the dissipation for the rough and smooth walls are also very close within the roughness layer. It means that, in terms of modeling of turbulent flows in rough pipes, a turbulence model based on smooth walls will suffice if an additional production term (related to the forcing on the roughness elements) is added within the roughness layer to produce an additional dispersive stress. 

The non-dimensional turbulent shear rate parameter, $S^*=P/\varepsilon$, representing the ratio of production of turbulent kinetic energy $P$ to its dissipation $\varepsilon$ is presented in figure~\ref{fig:stat20}. Although production is higher very close to the average wall, we also see that the peak value of the turbulent shear rate is lower for the rough wall. In the outer layer the curves for the smooth and rough walls merge as expected.
\subsubsection{Stresses anisotropy}
\begin{figure}
\centering
\includegraphics[width=0.55\textwidth]{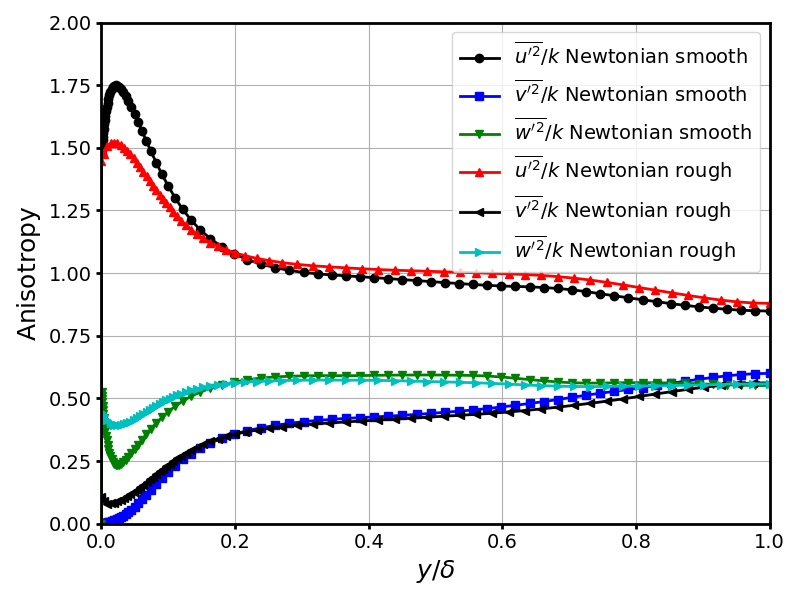}
\caption[]{Turbulence anisotropy in smooth wall versus rough wall conditions.}
\label{fig:stat10}
\end{figure}

Wall roughness increases the turbulence intensity in a finite roughness sublayer (the near-wall region that is affected by the roughness, and postulated to exist for $\delta/r > 40$). The effect of roughness on the degree of turbulence anisotropy can be evaluated by the ratio of the individual normal turbulent stress components to the turbulent kinetic energy ($k$), $\overline {u_iu_i}/k$, as shown in figure~\ref{fig:stat10}. It shows that the anisotropy is significantly reduced near the wall. The contribution of the streamwise fluctuation to the total turbulent kinetic energy is significanly lower. From $y/\delta > 0.1$ the anisotropy remains practically unchanged. In the channel center, the spanwise and vertical velocity fluctuations reach similar values showing cross-stream homogeneity.
\subsection{Wall friction}
\label{wallfriction}
\begin{figure}
\center
\includegraphics[scale=0.2,angle=0]{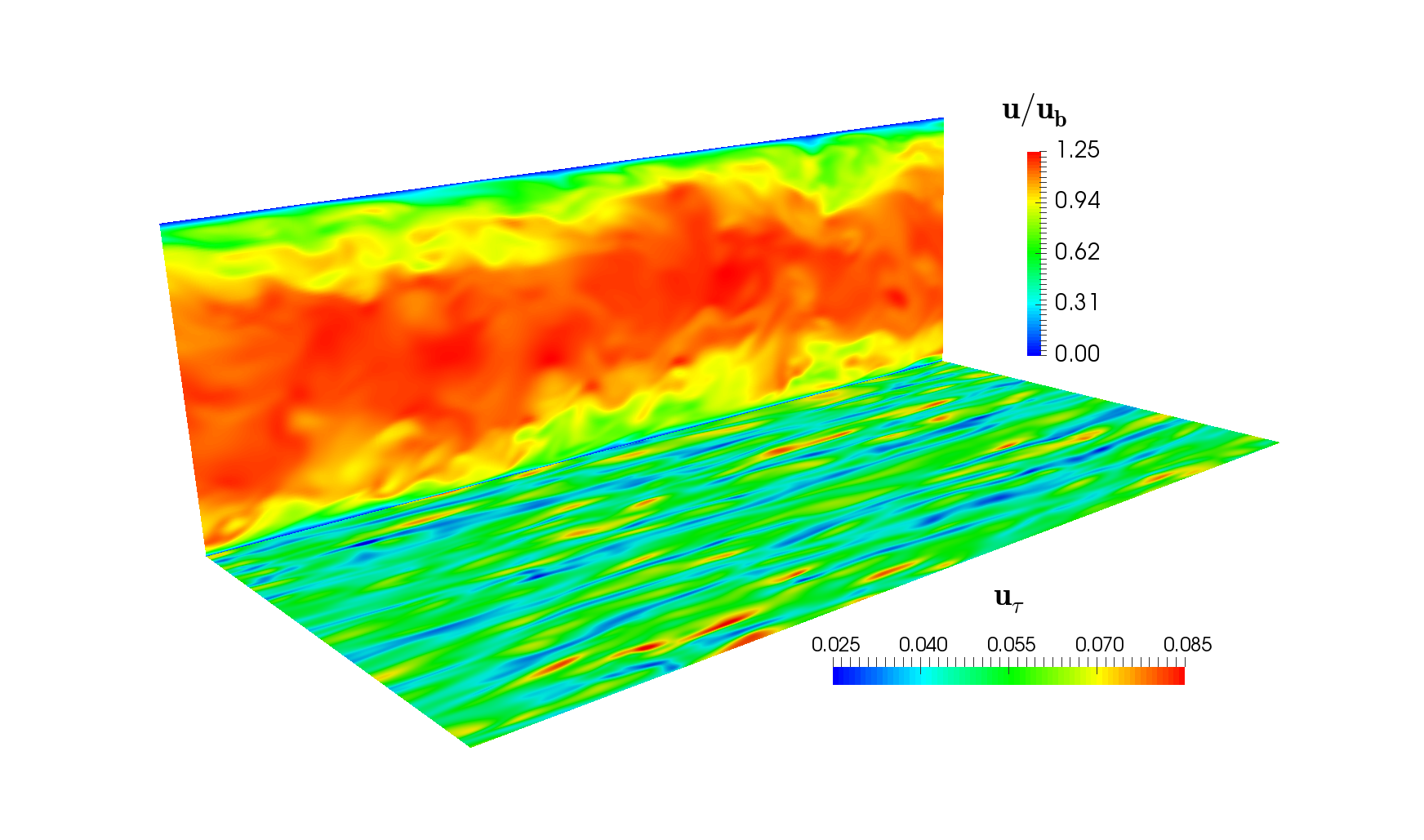}
\includegraphics[scale=0.2,angle=0]{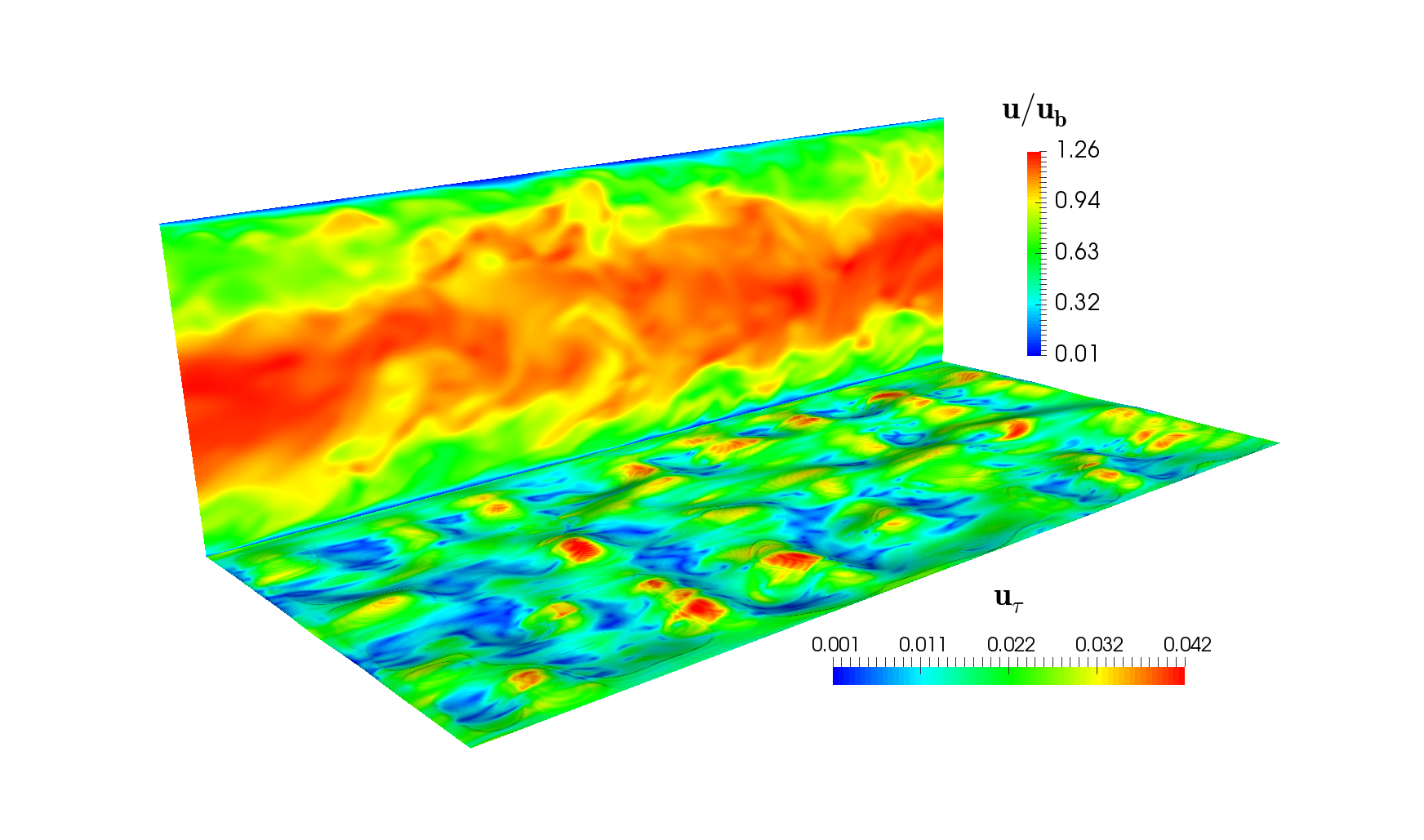}
\caption[]{Fluctuating field and shear velocity at the wall: Smooth vs. rough wall results.}
\label{results:NSrecirc}
\end{figure}

Figure~\ref{results:NSrecirc} provides a comprehensive idea about the structure of turbulence as it evolves in the flow direction in each case. In addition to the contours of instantaneous velocity, the figure embeds wall-contours of the friction velocity. The smooth-wall results clearly show the streaky structure pattern, which consists of alternating regions of low- and high-speed fluid.
The case with roughness suggests that these near-wall streaks are now considerably shortened due to the crests of the surface. The streamwise correlation lengths in the wall region are now smaller and also proportional to the roughness wavelengths. 
As shown by \cite{ma_alamé_mahesh_2021} the crests of the undulating wall surface are associated with higher shear stress regions, and the troughs/valleys with low-shear stress where reverse flow occurs. The same phenomenon was incidentally observed in the DNS of interfacial, sheared air-water flow of \cite{fulgosi:jfm}. Further, the coherence of the streaky structures over the rough wall is affected, and the flow establishes in the new patterns sketched by the roughness surface.

The friction and bulk velocities, and the corresponding Reynolds numbers predicted by the Colebrook-White equation \citep{menon2014transmission} are presented in table~\ref{colebrookwhitecomparison}. For the smooth wall, the DNS results are within $3$\% of the correlation. For the rough wall, the friction factor was calculated by specifying the roughness to be $7.5$ wall units. In this case $u_{\tau}$ is predicted to within $2$\% and the bulk velocity within $9$\% (as with the corresponding Reynolds numbers). This points to the fact that the roughness characteristics of the walls used in this study are of a generic nature with an expectation to satisfy well established correlations. 
\begin{table}
\center
\begin{tabular}{ccccccccc}
\hline
 & \multicolumn{4}{c}{DNS Results} & \multicolumn{4}{c}{Colebrooke-White} \\ \hline
\makebox[1.1cm]{\textbf{Case}} 
& \makebox[1.1cm]{$u_{\tau}$} & \makebox[1.1cm]{$u_b$} &  \makebox[1.1cm]{$Re_{\tau}$} & \makebox[1.1cm]{$Re_b$} & \makebox[1.1cm]{$u_{\tau}$} & \makebox[1.1cm]{$u_b$} &  \makebox[1.1cm]{$Re_{\tau}$} & \makebox[1.1cm]{$Re_b$}
\\ \hline
NS  & 0.0500 & 0.88 & 360 & 6363 & 0.0500 & 0.908 & 360 & 6538 \\ \hline
NR  & 0.0435 & 0.70 & 313 & 5039 & 0.0428 & 0.762 & 308 & 5486 \\ \hline
\end{tabular}
\caption[]{Comparison between DNS and Colebrook-White predictions}
\label{colebrookwhitecomparison}
\end{table}

The reduction of the bulk velocity due to roughness because of form drag leads to a significant increase in the friction factor (Darcy-Wiesbach friction factor, defined as $f = 8 (u_*/U_b)^2$, where $U_b$ is the flow bulk velocity) as shown in table~\ref{table:frictionN}. A 58\% increase in the friction factor is observed for the rough case, which according to the definition of the friction factor can be explained by a reduction in the flow rate or bulk velocity for a specified pressure gradient ($u_{*}$ is the same). This increase in the friction corresponds to a reduction of the flow rate by 20\% because of roughness. The Darcy-Wiesbach friction factor estimated using the Colebrook-White equation is also presented in table~\ref{table:frictionN}. The friction factor predicted for rough walls using the correlation is 15\% lower than the DNS result mainly due to the error in the bulk velocity that gets amplified in the friction factor due to squaring.
\begin{table}
\center
\begin{tabular}{ccc}
\hline
{\textbf{Case}} & {\textbf{Friction factor}} & {\textbf{Colebrook-White}} \\ \hline
\textbf{NS} & $0.0258$ & 0.0243 \\ \hline
\textbf{NR}  & $0.0408$ & 0.0345 \\ \hline
\end{tabular}
\caption{Newtonian friction factor}
\label{table:frictionN}
\end{table}
\subsection{Flow structures}
\begin{figure}
\includegraphics[scale=0.12,angle=0]{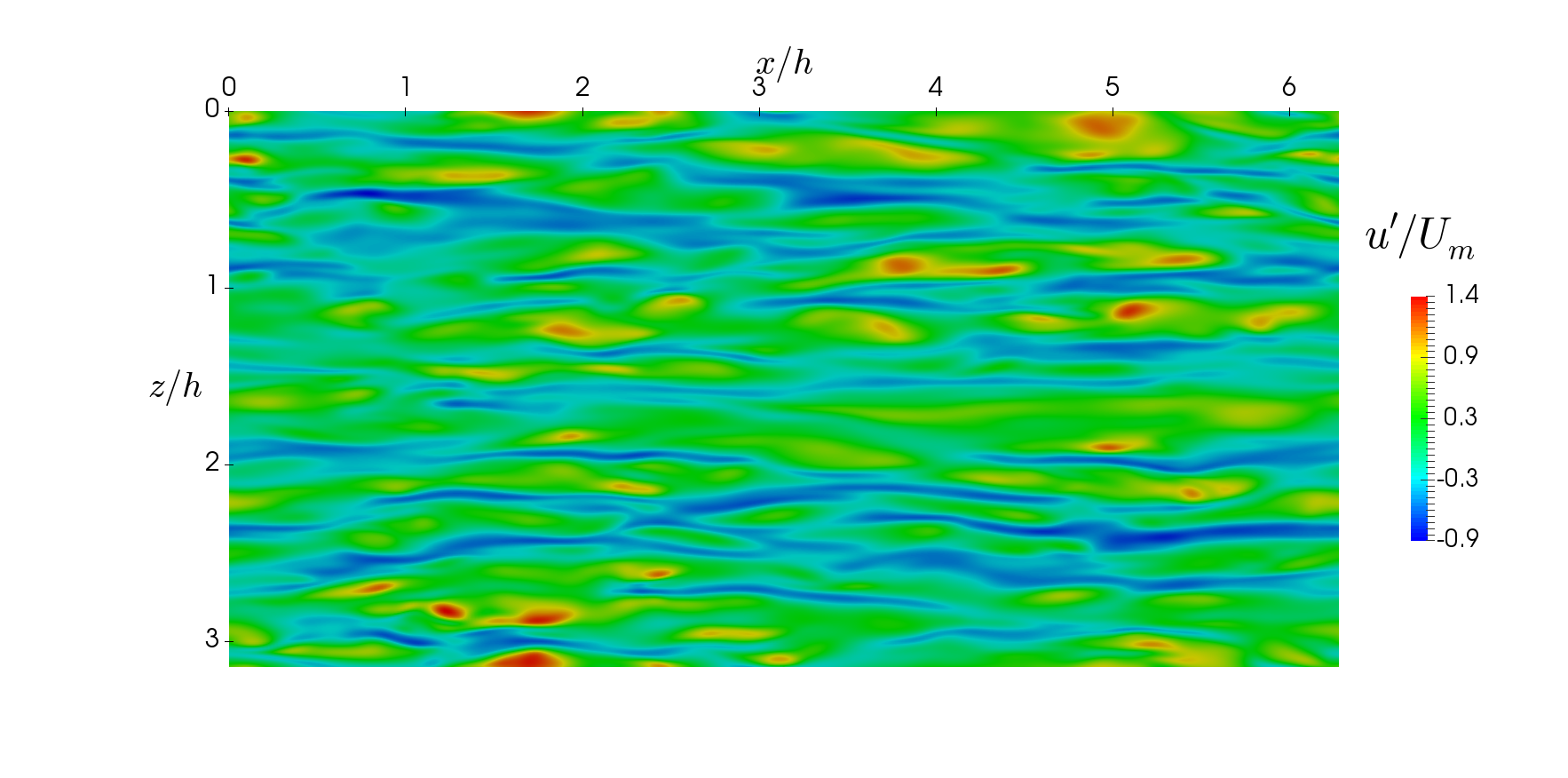} \hspace{-0.5cm} 
\includegraphics[scale=0.12,angle=0]{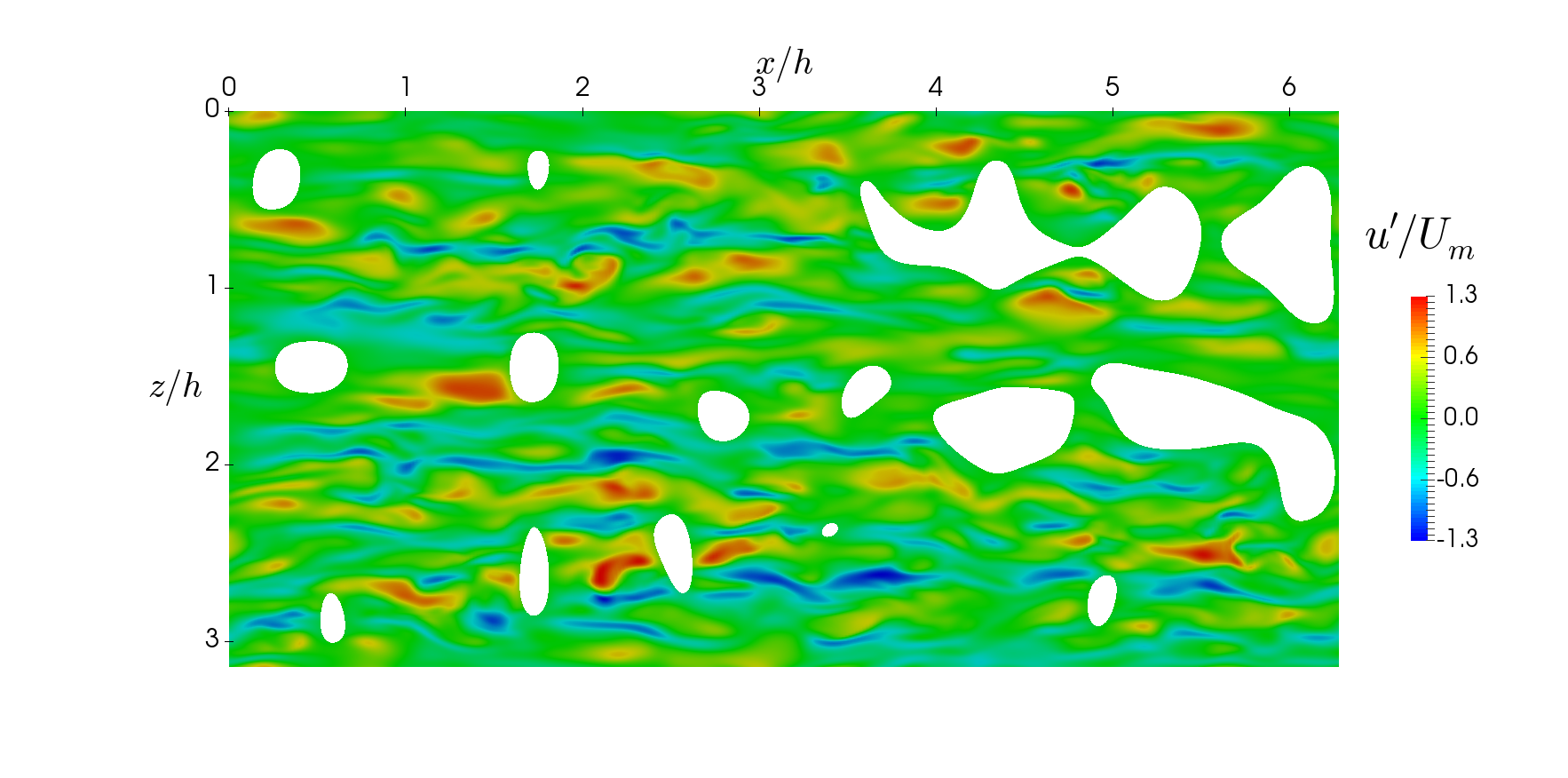} 
\hspace{-0.7cm} 
\includegraphics[scale=0.12,angle=0]{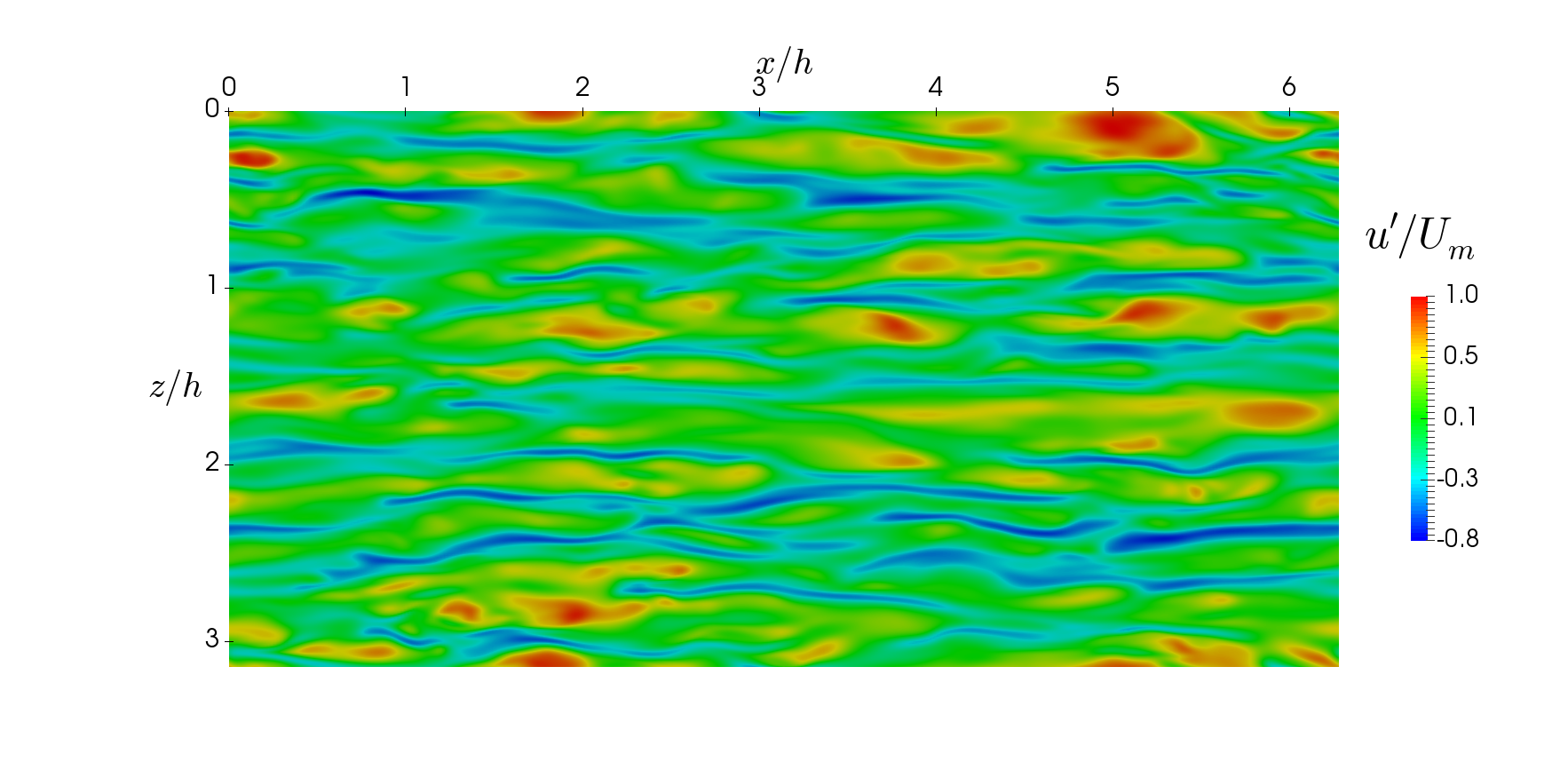} \hspace{-0.5cm}
\includegraphics[scale=0.12,angle=0]{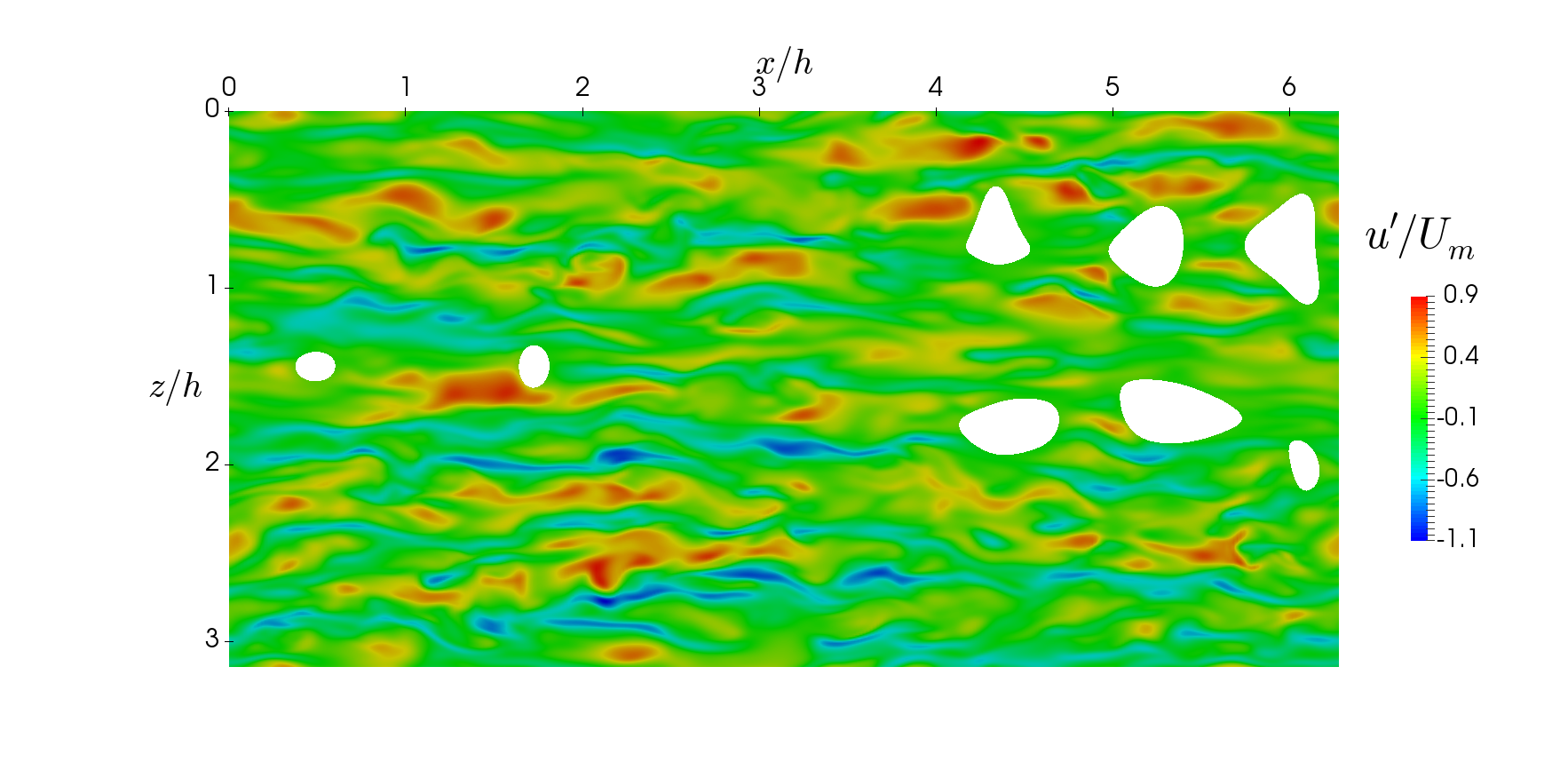}
\hspace{-0.7cm} \includegraphics[scale=0.12,angle=0]{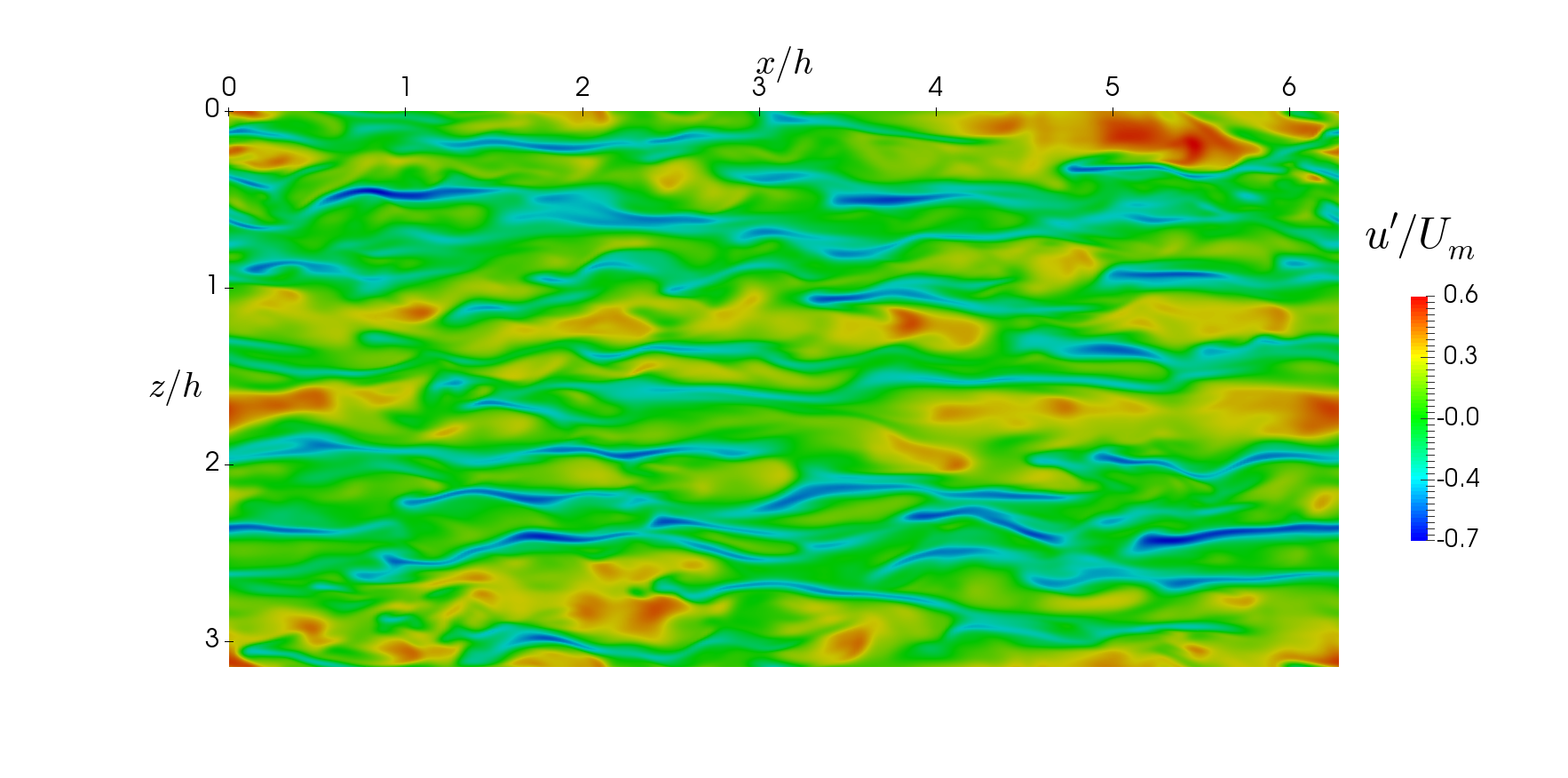} \hspace{-0.5cm}
\includegraphics[scale=0.12,angle=0]{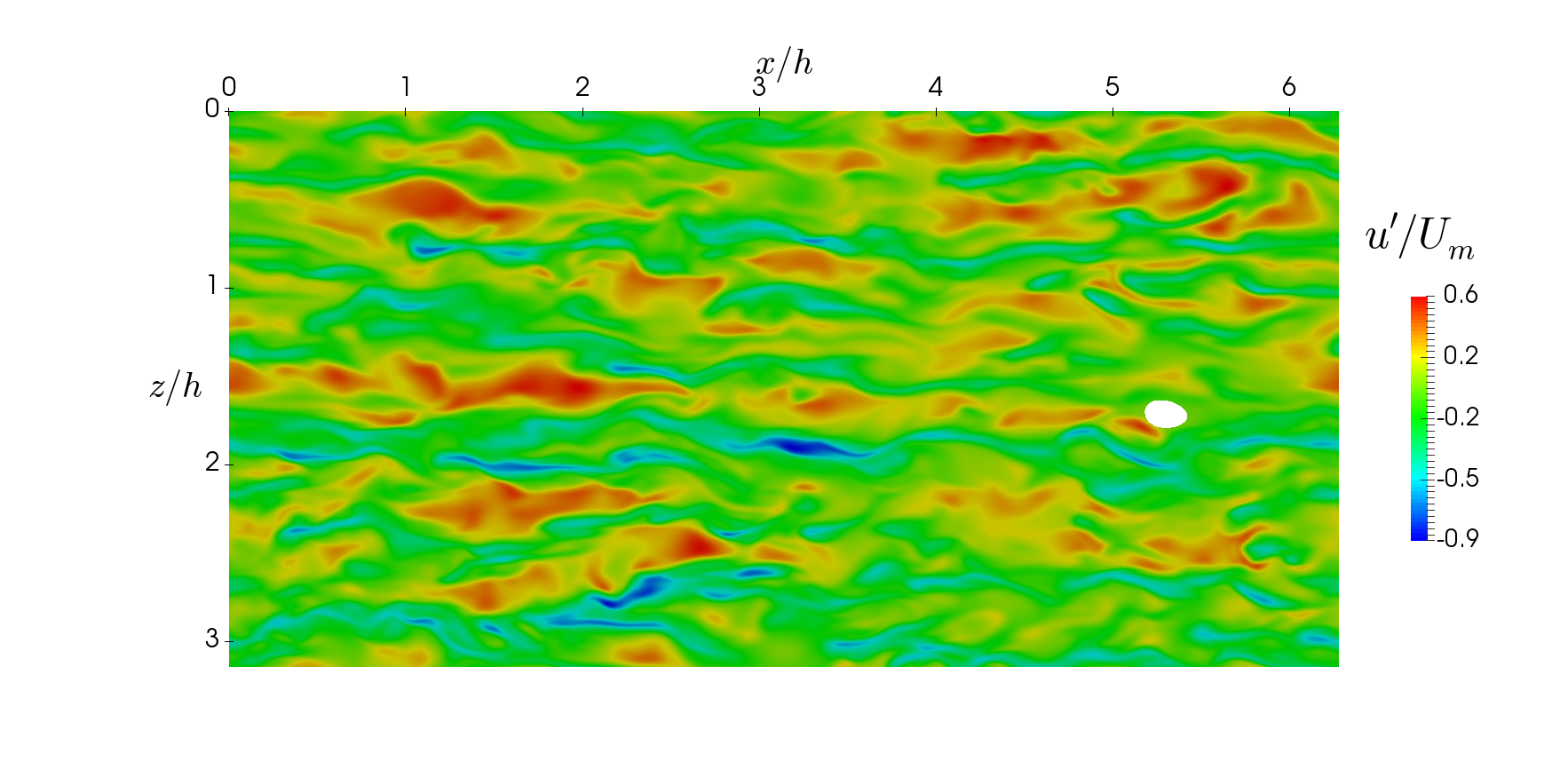}
\hspace{-0.7cm} \includegraphics[scale=0.12,angle=0]{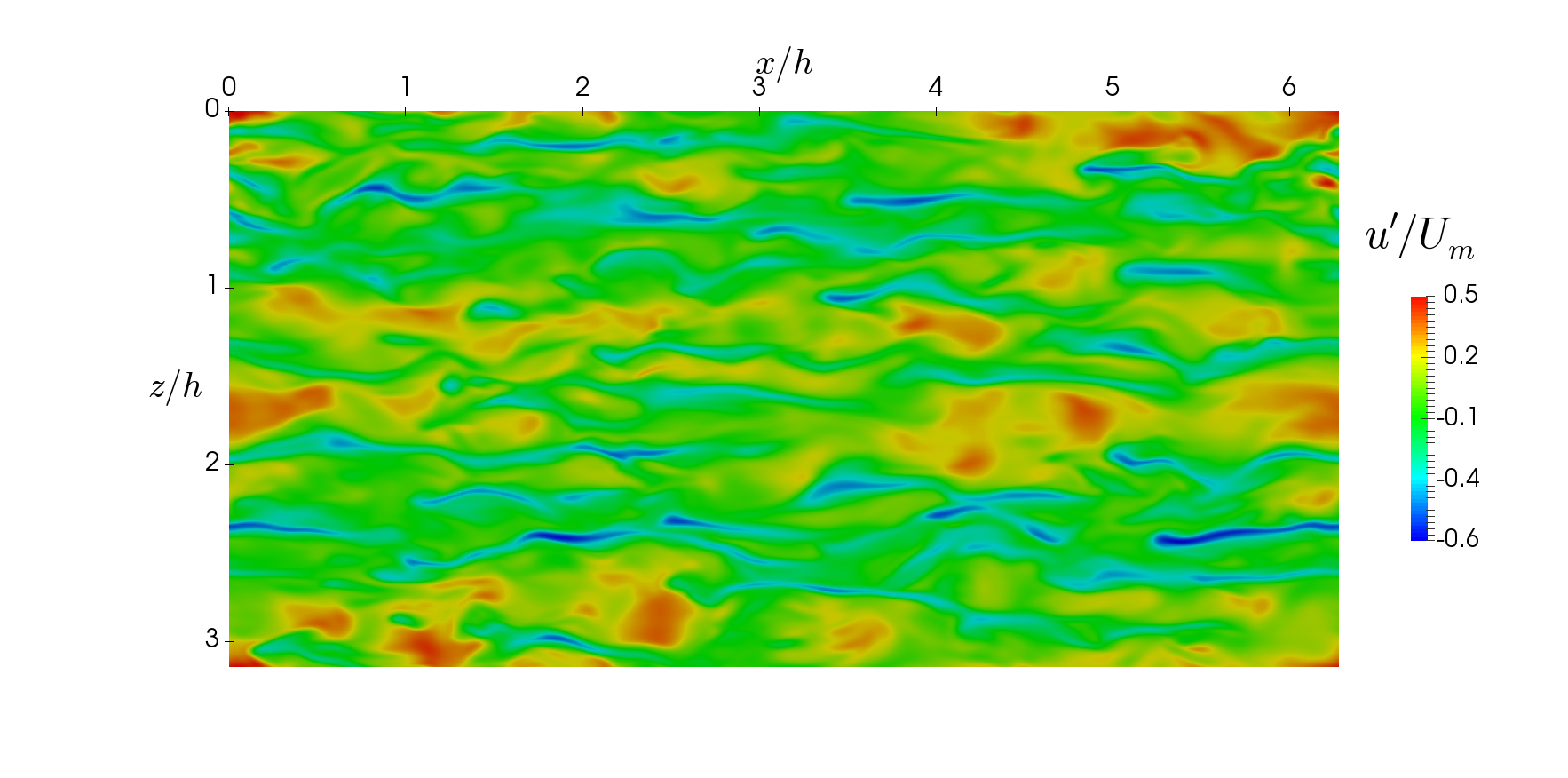}
 \hspace{-0.5cm}\includegraphics[scale=0.12,angle=0]{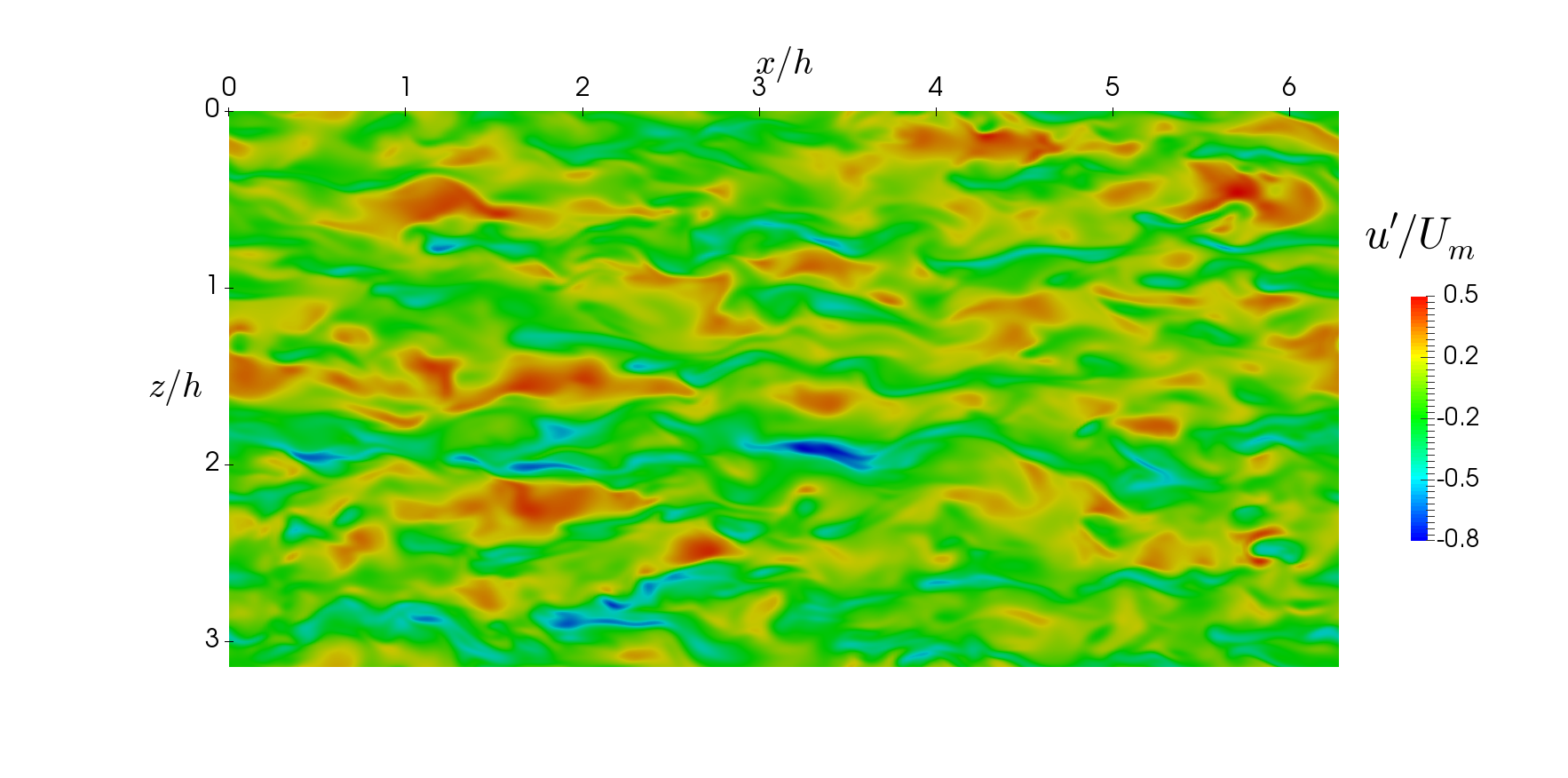}
\caption[]{Instantaneous axial flow velocity in the smooth channel at $y^+ = 5, 10, 20, 30$ for Newtonian fluid.}
\label{results:NSU}
\end{figure}

The effect of roughness on the flow in the wall layer is brought by looking at the fluctuating streamwise velocity contours. This quantity is displayed in figure~\ref{results:NSU} at four near-wall $x-z$ planes: $y^+ = 5, 10, 20, 30$. The instantaneous flow field is normalized by the bulk velocity $u_b$. The marked differences that could be observed between smooth- and rough-wall simulation cases featuring large obstructions are not seen here, albeit the structures seem to be slighted lifted upward by almost a constant wall-unit shift of $y^+ = 5$. 

The streaky structures are clearly visible in the viscous-affected layer (i.g. the closest $x-z$ planes to the wall), for both smooth and rough cases. Two main differences can be highlighted. First, in the smooth case the streaky structures are elongated with the flow, sometimes occupying the entire domain. In contrast, in the rough case, the structures are broken by the roughness elements. Long structures are not given sufficient time to develop. As a consequence of this blockage effect, the dissipation mechanisms induced
by vortex stretching are reduced. While the roughness protrusions tend to disrupt the formation of long structures, the effect is clearly less pronounced than in \cite{busse2017reynolds}, or in the case where the boundary layer is affected by air bubbles acting as roughness asperities \citep{lakehal2017turbulent} of a similar scale. Secondly, the smooth case panels clearly show more low-momentum regions as compared to the rough-wall results. Further, in average, one could speculate visually that the flow is overall faster in the roughness layer, conforming the observation made in the context of figure~\ref{results:newtoniancomparison}. The effect of roughness suddenly vanishes at $y^+ = 30$, which could thus be marked as the critical roughness height for the outer-layer similarity. For the type and characteristics of roughness considered here, the layer directly influenced by the roughness is confined to a region of $20 < y^+ < 25$ from the wall.

The population of a turbulent flow with streaky structures can as well be evaluated through the turbulent shear rate parameter, $S^*=P/\varepsilon$. For $S^* > 1$, the shear is strong enough for streaks to form, indicating that the generation of turbulence is more dominant than its dissipation. Looking back at figure~\ref{fig:stat20}, comparing this quantity for smooth and rough-surface flows, indicate that the streaks form at almost the same distance from the wall, being more pronounced in the smooth case though, in line with the previous analysis of the near-wall fluctuating flow field. For the rough-wall flow the ratio does not decay quickly to zero at $y=0$ with a value of $\approx 1.25$ at the average wall location, which proves that the roughness layer is characterized by a denser population of coherent structures.
\subsection{Coherent structures}
\label{sec:visu}
In turbulent flow, the separation between the coherent and non-coherent motion helps quantify the process of energy production and transfer between the mean and the fluctuating field, and among the turbulent stress components \citep{JeongHussain}. Coherent structures (CS) consist of streaks and streamwise vortices, linked to ejections and sweeps. These are responsible for draining the slow-moving fluid into the outer region and the high-momentum fluid towards the wall. In addition, these events generate the major part of the drag and should correlate with heat and mass transfer fluxes. In the present context, the analysis should also tell whether there exists a direct correlation between the CS population in the roughness layer and in the outer layer. 

There exists several sophisticated eduction techniques to characterize qualitatively the quasi-streamwise vortices in turbulent flow. Here, we rely on the so-called Q vortex identification criterion \citep{hunt}, a well known measure of the balance between the rate-of-rotation tensor $r_{ij}$ and the rate-of-strain tensor $s_{ij}$ within the superimposed fluctuating field, and is defined as $Q = 1/2(r_{ij}r_{ij} - s_{ij} s_{ij})$; the positive values of which indicate regions where the strength of the rotation overcomes the strain. For this purpose, the probability density functions (PDF) of the identifier has been determined and, the isosurface thresholds of PDF $0.005$ was selected. 

\begin{figure}
\center
\includegraphics[width = 0.8\textwidth]{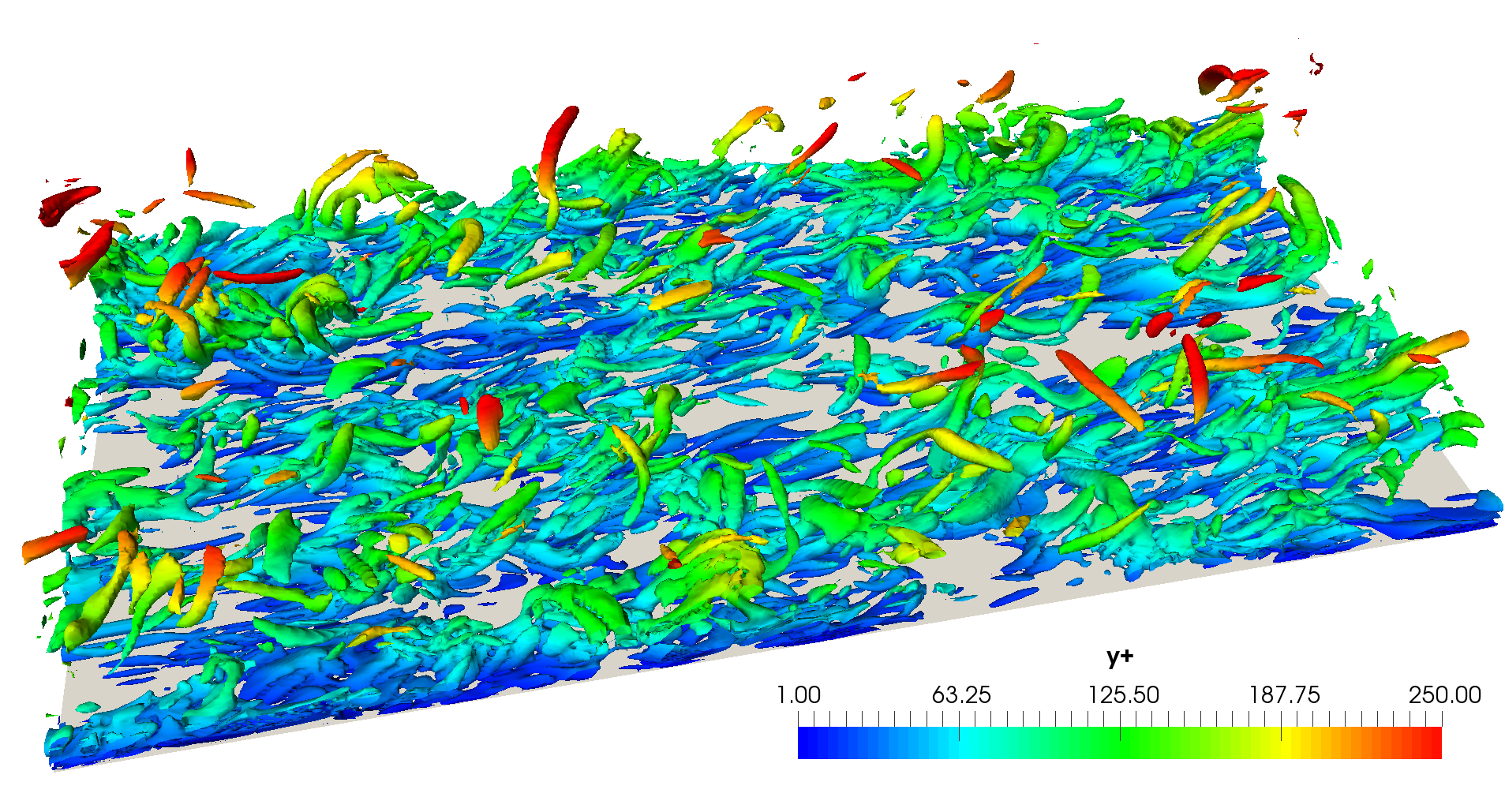}
\includegraphics[width = 0.8\textwidth]{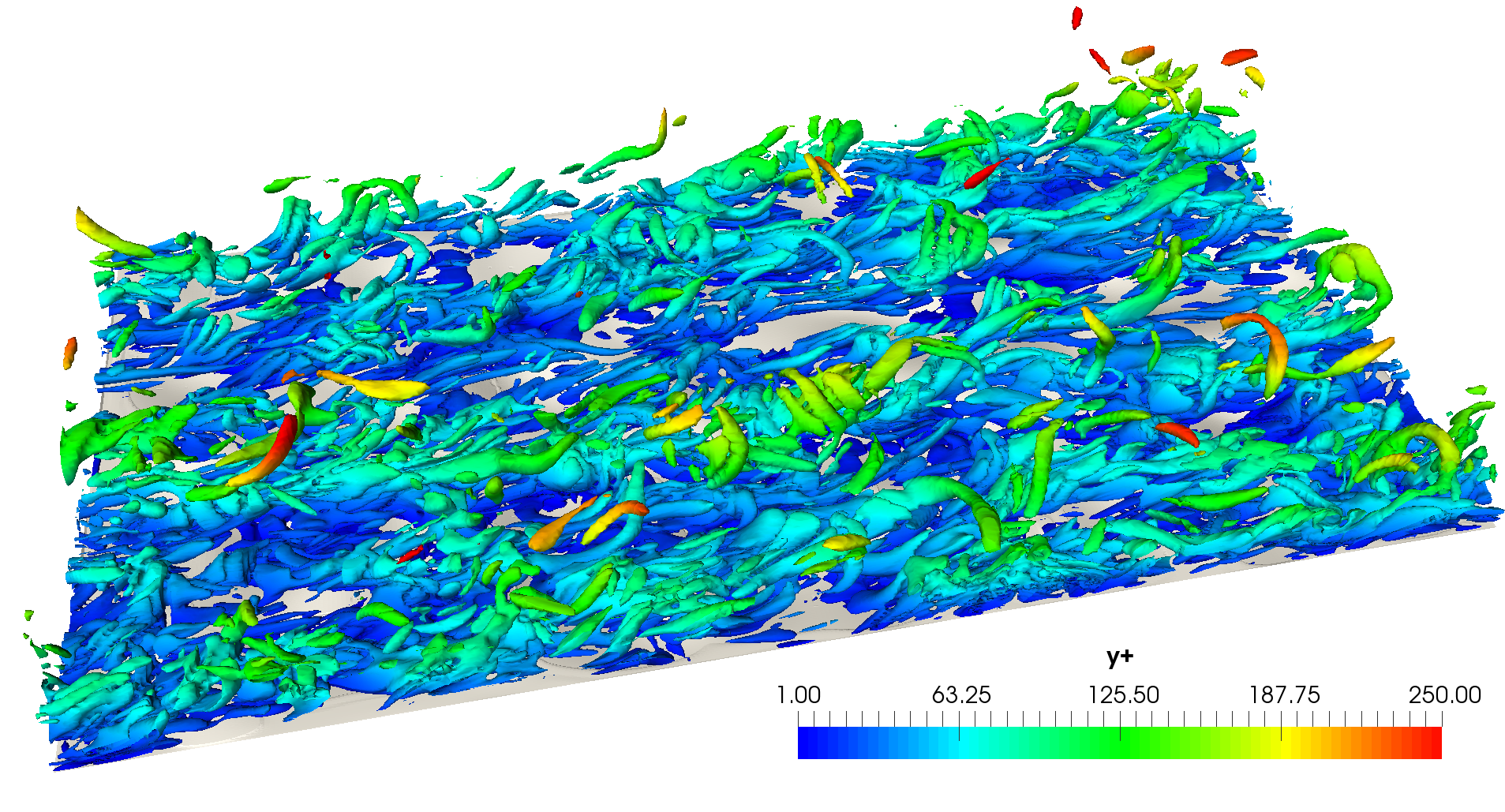}
\caption[]{Contour of the Q-criterion with $\textrm{PDF} = 0.005$, smooth vs. rough wall.}
\label{fig:q_criterion_yplus1}
\end{figure}

3D views of the vortex identification criterion are depicted in figure~\ref{fig:q_criterion_yplus1} for smooth- and rough-wall flows. The instantaneous isocontours (taken at a randomly-selected simulation instant) of $Q_{PDF=0.005}$ colored by $y^+$ show the nature of the structures and their concentration regions : rather locally small-scale vortices surrounded by hairpin-like structures. Obviously, because it is based on velocity gradients, the Q criterion cannot return the large scale motions, which can only be characterized by the velocity itself. Independent of the wall surface, the structures are either elongated in the streamwise direction or inclined with a some inclination angle. The view confirms the results discussed earlier, that is, in the roughness layer, the vortical structures are broken and disrupted by the surface elements. Further, the structures do not seem to rise towards the outer layer as in the case of large, cube-type of roughness shown by many authors. Likewise, the interaction between the flow and the roughness elements is not so obvious, although the lift of the flow due to their blockage effect is perceivable. 
A part of this, the results raise two instructive findings: surface roughness generates a denser population of CS, confined in the affected layer $y^+<25$, than in the smooth case, and the hypothetical penetration of these structures to the outer layer is not evidenced. On the contrary, it rather seems the outer layer has way less structures than the smooth case, which corroborates with the 
 observation concerning $S^*$, made in the context of figure~\ref{fig:stat20}.
This observation strengthens the conclusion drawn so far as to no direct impact of the roughness layer on the outer one. This is said, other methods do exist for a detailed quantitative description of the structures, which should help better assess the similarity of the flow structures appearing in the near-wall region and outer region \citep{Wang2021}.
\section{Conclusions}
DNS of turbulent flow in a rough channel have been performed for a shear Reynolds number of $Re_{*}=360$ with the objective of characterizing the effects of surface roughness on non-Newtonian fluid flow. The rough surfaces studied were made of smooth irregular protrusions and differ from the sharp-edged, bar-type and cube-type of roughness investigated in many previous studies. The maximum crest and trough heights were 23 wall units and the average height was found to be 5 wall units, with a 100 wall-unit spacing. The RMS of the wall roughness, taken here as the proper roughness characteristic length scale $r_s$ is equal to $7.5$. The ratio of the channel half-height to the mean roughness and to its RMS is $71$ and $56$, respectively.

The dynamic force balance shows that the viscous shear at the wall accounts for $75\%$ of the applied pressure forcing and the remaining $25\%$ by form drag. A very good match to the modified log law could be obtained by using smooth-wall values of $\kappa = 0.41, B = 5.5$, and $r^+ = 7.5$. This result, obtained only by means of a generic roughness adjustment, indicates that the roughness characteristics selected are of a generic nature and therefore appropriate for the study of roughness effects on Newtonian and non-Newtonian flows in pipelines. The smoothness of the roughness elements could be a reason to expect universal behaviour that has evaded earlier studies. This is further supported by the fact that the present data reaffirms the hypothesis that the ratio of form drag to the total drag depends only on the effective slope of the rough surface. 

In addition, the structure of the turbulent boundary layer with regard to satisfying a log law or a power law has been analyzed by plotting $\gamma$ and $\beta$ profile characteristics parameters. For the rough surface, the interval where a log-law behaviour can be expected is smaller with a lower effective $\kappa$. On the other hand, the variation of $\beta$ shows that for both the smooth and rough cases, the mean velocity profile could be very well represented as a power law.

The results show that when normalized by the friction velocity $u_{\tau}$, all turbulent stresses are significantly higher in the roughness sublayer ($y^+ < 25$), exceeding slightly the peak roughness height ($\approx 23$ wall units). The turbulent shear stress $\overline{u'v'}$, in particular, shows the largest increase due to roughness. Production of turbulence is promoted due to increase in the turbulent stress even though there is a reduction in the mean velocity gradient. When using $u_*$ the effect of roughness in the outer region is revealed, notably, the normal stresses match the smooth wall almost exactly. This shows that for the structures height studied, the effect of the roughness is not perceived in the outer region consistent with the observations of several earlier studies. In other words, for a given driving force, increased turbulence production and intensity is restricted to the inner region only. In summary, except in the roughness sublayer, the Reynolds stresses for the rough surface collapse with smooth-wall results.

The mean shear stress balance in the rough channel shows that the total stress peaks away the average wall. Close to the wall, the viscous shear is of the same magnitude as the turbulent stress thereby shifting the point at which the two stress components are in balance closer to the wall. In the outer scaling the effect of roughness is not visible beyond $y/\delta = 0.1$.

The turbulent kinetic energy budget shows that turbulence production is significantly higher in the viscous layer for rough walls such that it dominates the mean viscous diffusion. Turbulent dissipation, therefore, balances the sum of mean viscous diffusion and turbulence production. The effect of roughness is likewise noticeable in the distribution of the turbulent stresses and the inter-stress energy transfer. Roughness significantly reduces stress anisotropy near the wall. The contribution of the streamwise fluctuation to the total turbulent kinetic energy in the roughness sublayer is significantly lower for the rough case. From $y/\delta= 0.1$ the anisotropy remains practically unchanged. In the channel center, the spanwise and vertical velocity fluctuations reach similar values, showing cross-stream homogeneity. 

In relation to the friction factor, the DNS results are within $3\%$ of the Colebrook-White correlation for the smooth wall. For the rough wall case, the friction factor calculated by specifying the roughness characteristic scale $r_s$ to be 7.5 wall units, predicts the friction velocity $u_{\tau}$ to within $2\%$ and the bulk velocity within $9\%$ (as with the corresponding Reynolds numbers). This reaffirms the statement that the roughness characteristics of the walls in this study are of a generic nature thereby satisfying well established correlations. The reduction in the flow rate or bulk velocity for a specified pressure gradient because of form drag leads to a significant increase in the friction factor. In the present case, a $58\%$ increase in the friction factor is observed. This increase corresponds to a reduction of the flow rate by about $20\%$. Form drag is produced mainly by the roughness protrusions, forcing the mean-flow momentum to drift downward by the viscous force to balance the total drag induced by the roughness.

The modifications of the near-wall flow structure due to roughness were quantified by exploring the fluctuating flow fields. In the smooth case the streaky structures are elongated along the flow, sometimes occupying the entire domain, in contrast, for the rough case the structures are broken by the roughness elements, thereby preventing the development of long structures. This blockage effect of the flow field creates additional flow interactions in the near-wall region, inducing larger velocity fluctuations. The alterations of the coherent structures by roughness were examined through the analysis of the instantaneous flow fields and their statistical properties. The analysis raises two instructive findings : the roughness sublayer is covered by a denser population of coherent structures than in the smooth case, and no facts could be evidenced as to the direct impact of the roughness layer on the outer one.
\newpage
\bibliographystyle{jfm}
\bibliography{jfm}
\end{document}